\def\b{\bibitem}
\def\be{\begin{equation}}
\def\ee{\end{equation}}
\def\bea{\begin{eqnarray}}
\def\eea{\end{eqnarray}}
\def\bml{\begin{mathletters}}
\def\eml{\end{mathletters}}
\documentstyle[aps,prb,eqsecnum,psfig,epsf,floats]{revtex}
\draft
\begin{document}
% Macros for the various macro package names, etc.
\def\SNG{{\em Physical Review Style and Notation Guide}}
\def\LUG {{\em \LaTeX{} User's Guide \& Reference Manual}}
\def\btt#1{{\tt$\backslash$\string#1}}%
\def\REVTeX{REV\TeX}
\def\AmS{{\protect\the\textfont2
        A\kern-.1667em\lower.5ex\hbox{M}\kern-.125emS}}
\def\AmSLaTeX{\AmS-\LaTeX}
\def\BibTeX{\rm B{\sc ib}\TeX}
%\makeatletter
%\tighten
\twocolumn[\hsize\textwidth\columnwidth\hsize\csname@twocolumnfalse%
\endcsname
\title{Long-Time Tails, Weak Localization, and Classical and Quantum Critical 
       Behavior\\
      % \small{$[$ J. Stat. Phys. {\bf xx}, xxxxx (2002) $]$}
                }
\author{T.R. Kirkpatrick$^{1,2}$, D. Belitz$^3$, and J.V. Sengers$^{1,4}$}
\address{$^1$Institute for Physical Science and Technology, University of 
             Maryland, College Park, MD 20742}
\address{$^2$Department of Physics, University of 
             Maryland, College Park, MD 20742}
\address{$^3$Department of Physics and Materials Science Institute,
             University of Oregon, Eugene, OR 97403}
\address{$^4$Department of Chemical Engineering, University of
             Maryland, College Park, MD 20742}
\date{\today}
\maketitle

\begin{abstract}
An overview is given of the long-time and long-distance behavior of
correlation functions in both classical and quantum statistical mechanics.
After a simple derivation of the classical long-time tails in equilibrium
time correlation functions, we discuss analogous long-distance phenomena
in nonequilibrium classical systems. The paper then draws analogies
between these phenomena and similar effects in quantum statistical
mechanics, with emphasis on the soft modes that underly long-time tails
and related phenomena. We also elucidate the interplay between critical
phenomena and long-time tails, using the classical liquid-gas critical
point and the quantum ferromagnetic transition as examples.
\end{abstract}
\pacs{KEY WORDS: Long-time tails; generic scale invariance; weak localization;
                 critical behavior; quantum phase transitions}
\vskip 20pt
]
%\narrowtext
\section{Introduction}
\label{sec:I}

\begin{figure}[ht]
\epsfxsize=47mm
\centerline{\epsffile{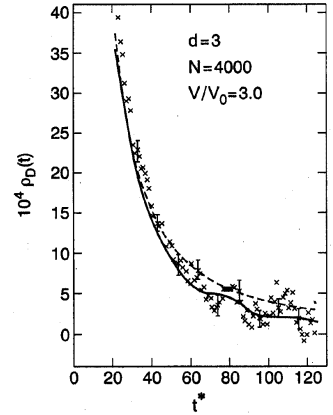}}
\vskip 5mm
\caption{Normalized velocity autocorrelation function $\rho_D(t) =
 \langle{\bf v}(t)\cdot{\bf v}(0)\rangle/\langle{\bf v}^2(0)\rangle$ as
 a function of the dimensionless time $t^* = t/t_0$, where $t_0$ is the
 mean-free time. The crosses indidate computer results obtained by
 Wood and Erpenbeck, Ref.\ \protect\onlinecite{WoodErpenbeck}, for a 
 system of $4000$ hard spheres at a reduced density corresponding to
 $V/V_0 = 3$, where $V$ is the actual volume and $V_0$ is the close-packing
 volume. The dashed curve represents the theoretical curve
 $\rho_D(t) = \alpha_D\,(t^*)^{-3/2}$. The solid curve represents a more
 complete evaluation of the mode-coupling formula with contributions from
 all possible hydrodynamic modes and with finite-size corrections
 included, Ref.\ \protect\onlinecite{Dorfman}. 
 From Ref.\ \protect\onlinecite{DKS}.}
 \vskip 0mm
\label{fig:1}
\end{figure}

It has been known for some time that equilibrium time correlation functions
generically show a power-law temporal decay in the limit of long times.
If this behavior is not obviously related to any conservation law, or to
soft modes in the system, long-time tails (LTTs) is the term traditionally
applied to this phenomenon.
Boltzmann-type kinetic equations usually predict an exponential decay in
time, which is why the discovery of LTTs came as a considerable surprise.
An example is the velocity
auto-correlation function in a hard-sphere fluid, which is shown in 
Fig.\ \ref{fig:1}.
In accord with the Boltzmann-Enskog equation, initially the decay is
exponential, with a time scale given by the mean-free time $t_0$
between particle collisions. However, for longer times, $t>>t_0$, it decays
algebraically, as $t^{-3/2}$, and more generally as $t^{-d/2}$ in
$d$-dimensions. This striking result was first observed in computer
simulations by Alder and Wainwright,\cite{AlderWainwright} 
and then understood theoretically by Dorfman and Cohen,\cite{DorfmanCohen}
and by Ernst, Hauge, and van Leeuwen.\cite{EHvL} 
The velocity autocorrelation function is of particular importance
because it determines the frequency-dependent self-diffusion
coefficient $D(\omega)$ via 
\begin{equation}
D(\omega) = \frac{1}{d}\int_{0}^{\infty }dt\ e^{i\omega t}\ 
            \langle{\bf v}(t)\cdot {\bf v}(0)\rangle\quad.  
\label{eq:1.1}
\end{equation}
Here ${\bf v}(t)$ is the velocity of a tagged particle at time $t$, and the
angular brackets denote an equilibrium ensemble average. The slow algebraic
decay of the velocity autocorrelation function implies a low-frequency
behavior
\bml
\label{eqs:1.2}
\bea
D(\omega\rightarrow 0)/D(0)&=&1 - c^{D}_{d}\,(i\omega t_0)^{(d-2)/2} + \ldots
\nonumber\\
&&\hskip 90pt (2<d<4)\quad,
\nonumber\\
\label{eq:(1.2a)}\\
D(\omega\rightarrow 0)/D(0)&=&1 - c^{D}_2\,\ln (i\omega t_0) 
     + \ldots\quad (d=2) \quad,
\nonumber\\
\label{eq:1.2b}
\eea
\eml%
with $c^{D}_d>0$.
This result indicates that in $d\leq 2$ the zero-frequency diffusion
coefficient does not exist, which in turn implies that the ordinary
local hydrodynamic description of a fluid is not valid in low dimensions. 
This conclusion is indeed correct, as detailed investigations 
show.\cite{2d_hydrodynamics} Note that in this classical case transport
in low dimensions is faster than diffusive.
More generally, it is known that all the transport coefficients in a fluid
are given by time integrals over appropriate time correlation functions, and
that all these time correlation functions have similar LTTs.\cite{BoonYip}

In contrast to this behavior of time correlation functions, 
the spatial decay of
static correlations at large distances in classical systems in equilibrium
is generically indeed exponential. An exception is the behavior at critical
points. It was known long before the discovery
of LTTs that at these isolated points in the phase diagram correlations
decay algebraically in both time and space, i.e., like power-laws with
universal critical exponents.\cite{Stanley,Fisher,HohenbergHalperin}
Algebraic decay implies scale invariance, i.e., the
correlation functions are generalized homogeneous functions of space and
time. Correlations that exhibit scale invariance in an entire region of
the phase diagram, rather than just at isolated points, are said to
exhibit generic scale invariance (GSI).\cite{DKS,Nagel} 
In this language, the LTTs are an example of GSI in the time domain.

A natural question is what happens to the correlation functions that exhibit
GSI as a critical point is approached, and how these correlations 
influence the
critical behavior itself. We will see below that the mechanism that causes
the LTTs in the entire fluid phase typically becomes amplified near critical
points and leads to nontrivial (i.e., non-van Hove) critical dynamics, or
critical singularities in various transport coefficients. 
Violation of the van Hove picture for the slowing down of critical
fluctuations became apparent when the thermal conductivity of fluids
was found to diverge at the critical point, while the viscosity of
liquid mixtures exhibited a weak divergence at the consolute 
point.\cite{Sengers_etal} These singularities were later understood to be due
to mode-coupling effects by 
Fixman,\cite{Fixman} and then by Kadanoff and Swift\cite{KadanoffSwift} 
and by Kawasaki.\cite{Kawasaki} Here mode-coupling means the coupling of
products of slow hydrodynamic modes, or excitations, that are in some
sense orthogonal to single hydrodynamics modes, to the single modes
whose correlations form the correlation functions in question.
Historically, the LTTs were discovered after these
critical singularities were observed, and it turned out that they, too, are
caused by mode-coupling effects.\cite{EHvL}

In a classical system in a nonequilibrium steady state (NESS) the situation
is still more complicated. As we will discuss at the end of the first part
of the present paper, in this case the same
mode-coupling correlations that lead to the equilibrium LTTs and to the
critical singularities, also lead to GSI in spatial correlations.

In recent years there has been a separate but analogous development in the
study of electronic systems at zero 
temperature.\cite{analogy_footnote} In this case
the mode-coupling effects are stronger than in classical systems, leading 
to even more dramatic effects. For example, they lead to the localization 
of electrons in disordered two-dimensional systems, even for arbitrarily
weak disorder.\cite{gang_of_4} 
In addition, because of the coupling of the electron
dynamics to the various electronic correlation functions, and because of the
coupling between statics and dynamics that is inherent in quantum statistical
mechanics,\cite{NegeleOrland} it turns out that in zero-temperature systems,
GSI exists in both spatial and temporal correlations even in equilibrium. 
In disordered systems these many-body quantum phenomena are known as 
weak-localization effects,\cite{WL_footnote} since they can be considered 
precursors to an Anderson or Anderson-Mott metal-insulator 
transition,\cite{Anderson,Mott,us_R} but analogous, if slightly weaker,
effects also occur in clean quantum systems.
In the second part of this paper we explain these 
phenomena, making connections with the GSI in classical systems. We then
explore how these generic correlations influence, and are themselves modified
by, critical behavior at quantum phase transitions. The latter are
defined as phase transitions that occur at zero temperature and are triggered
by some non-thermal control parameter, like pressure or 
composition.\cite{Sachdev,Sondhi_etal,us_March} 
The metal-insulator transitions mentioned above
are examples of quantum phase transitions.

As noted above, in the classical case the basic source of the power-law decay 
characteristic of GSI is two-fold. First, soft or massless modes must
exist. Their softness can arise either from some basic underlying
conservation law, or they can be Goldstone modes, i.e., arise from a 
spontaneously broken continuous symmetry.
Second, the theory must in some sense be nonlinear, so that
these soft modes couple to the physical observables. It is this coupling
that is missing in simple theories, and this omission results in the
erroneous prediction of exponential decay.
In both classical and quantum systems such coupling
mechanisms exist, and in this way the GSI effects found in both systems are
related.

The above discussion illustrates that it is incorrect to view LTT effects,
as it is sometimes done, as only leading to small corrections to transport
coefficients, with no deeper importance. To the contrary, as we have pointed 
out, they are responsible for a myriad of effects. Not only do LTTs cause 
the leading frequency and, in the quantum case, temperature corrections to 
the transport coefficients, but there are numerous cases where their effects
are greatly magnified, and they dominate the physics both qualitatively and
quantitatively. Examples include, systems in reduced dimensionalities, systems 
near critical points, systems in Goldstone phases, and nonequilibrium systems.
One of the goals of this paper is to illustrate how these effects come about,
and how they are connected.

We will proceed as follows. In Sec. \ref{sec:II} we present the dynamical
equations that describe fluctuations in a classical fluid. We use these
equations to first describe the LTTs in a classical fluid, and then to
describe critical dynamics near a liquid-gas critical point. We conclude
this section by describing the GSI that occurs in spatial correlations in a
nonequilibrium fluid. In Sec. \ref{sec:III} 
we consider the weak-localization effects in disordered electron systems, 
which are quantum analogs of the classical LTTs. Because in quantum
statistical mechanics these dynamical correlations couple to the static
correlations, they lead to GSI in spatial correlations even in equilibrium
systems, as long as they are at zero temperature. We then discuss how this
GSI is important for describing quantum phase transitions such as the 
ferromagnetic transition in itinerant electron systems at zero temperature. 
This amounts to a discussion of how the
quantum LTTs affect quantum phase transitions. We
also discuss how nonequilibrium effects lead to even
longer-range correlations in quantum systems at zero temperature. We conclude
with a discussion of our results in Sec. \ref{sec:IV}.

\section{GSI and phase transitions in classical systems}
\label{sec:II}

\subsection{Fluctuating hydrodynamics}
\label{subsec:II.A}

To be specific, let us consider a classical fluid as an example of a
finite temperature, or classical, statistical mechanics system.
In general we are interested in the long-time and large-distance
behavior of systems, so we will always be concerned with the identification
of the relevant soft, or hydrodynamic, variables that determine this
behavior. In our case they are the conserved
variables of mass density $\rho $, momentum density ${\bf g} = \rho {\bf u}$,
with ${\bf u}$ the fluid velocity, and energy density $\epsilon$, and
the relevant low-frequency and long-wavelength dynamical equations are
the Navier-Stokes equations. To describe fluctuations, a Langevin force is
added so that the equations are\cite{fluctuating_hydrodynamics} 
\bml
\label{eqs:2.1}
\bea
\partial_t\,\rho + \nabla\cdot{\bf g}&=&0\quad,  
\label{eq:2.1a}\\
\partial_t\,g_{\alpha} + \nabla_{\beta}\,(g_{\alpha}u_{\beta})
  &=&-\nabla_{\alpha}\,p + \nabla_{\beta}\Bigl[\eta\,(\nabla_{\alpha}u_{\beta}
        + \nabla_{\beta}u_{\alpha}) 
\nonumber\\
&& \hskip -45pt +\,(\zeta - \frac{(d-1)}{d}\eta)\,
            \delta_{\alpha\beta}(\nabla\cdot{\bf u}) + P_{\alpha\beta}
                \Bigr]\quad,  
\label{eq:2.1b}\\
\rho\,T(\partial_t + {\bf u}\cdot\nabla)\,s&=&\nabla_{\alpha}(\lambda
              \nabla _{\alpha }T+q_{\alpha })  \quad.
\label{eq:2.1c}
\eea
\eml%
Here $p$ is the pressure, $T$ the temperature, $\eta$ the shear viscosity,
$\zeta$ the bulk viscosity, and $\lambda$ the thermal conductivity, and a
summation over repeated indices is implied. We have
chosen to use the entropy density $s$, rather than the energy $\epsilon$,
as our hydrodynamic variable, and in Eq.\ (\ref{eq:2.1c}) we have 
neglected a viscous
dissipation term that represents entropy production since it is
irrelevant to both the leading LTTs, and to the leading critical dynamics
singularities. The Langevin forces $P_{\alpha\beta}$ and $q_{\alpha}$
are uncorrelated with the initial velocity, and satisfy
\bml
\label{eqs:2.2}
\bea
\langle P_{\alpha\beta}({\bf x},t)\,P_{\mu\nu}({\bf x}',t')\rangle
  &=&2k_{\rm B} T\Bigl[\eta\,(\delta_{\alpha\mu}\delta_{\beta\nu} 
        + \delta_{\alpha\nu}\delta_{\beta\mu}) 
\nonumber\\
&& \hskip -65pt + (\zeta -\frac{(d-1)}{d}\eta)\,\delta_{\alpha\beta}
   \delta_{\mu\nu}\Bigr]\,\delta({\bf x}-{\bf x}')\,\delta (t-t') \quad,
\nonumber\\
\label{eq:2.2a}\\
\langle q_{\alpha}({\bf x},t)\,q_{\beta}({\bf x}',t')\rangle
&=&2k_{\rm B}\,\lambda\, T^2\,\delta_{\alpha\beta}\,\delta({\bf x}-{\bf x}')\,
     \delta (t-t')\ ,
\nonumber\\
\label{eq:2.2b}\\
\langle P_{\alpha\beta}({\bf x},t)\,q_{\mu}({\bf x}',t')\rangle&=&0\quad,
\label{eq:2.2c}
\eea
\eml%
where $k_{\rm B}$ is Boltzmann's constant. 
The above equations can be derived in a
number of ways and are known to exactly describe the long-wavelength and 
low-frequency fluctuations in a fluid.\cite{fluctuating_hydrodynamics}

\subsection{Long-time tails}
\label{subsec:II.B}

To illustrate the LTTs we choose a slightly different example than the
self-diffusion coefficient discussed in the Introduction, namely, the
shear viscosity of a classical fluid. We present a simplified calculation
of this transport coefficient, and then
discuss more general results. For simplicity, we assume an incompressible 
fluid. The mass conservation law is then trivial,
\be
\nabla\cdot {\bf u}({\bf x},t) = 0\quad,
\label{eq:2.3}
\ee
and the momentum-conservation law is a closed PDE for ${\bf u}$,
\be
\partial_t\,u_{\alpha} + ({\bf u}\cdot \nabla)\,u_{\alpha} = -\nabla_{\alpha}
 (p/\rho) + \nu \nabla^2 u_{\alpha} + \nabla_{\beta}P_{\alpha\beta}/\rho\quad.
\label{eq:2.4}
\ee
Here $\nu = \eta/\rho$ is the kinematic viscosity, which we assume to be
constant. The only role of the pressure term is to ensure that the
incompressibility condition is satisfied. In fact, it can be eliminated by
taking the curl of Eq.\ (\ref{eq:2.4}), which turns it into an equation for
the transverse velocity, see Ref.\ \onlinecite{Chandrasekhar} and below. 
The cause of the LTTs is the coupling 
of slow hydrodynamic modes due the nonlinear term in Eq.\ (\ref{eq:2.4}). 
We treat that term as a perturbation, i.e. we formally multiply it by
a coupling constant $\gamma$ (whose physical value is unity), and
calculate its effect to first order in $\gamma$.

\begin{figure}[htb]
\epsfxsize=40mm
\centerline{\epsffile{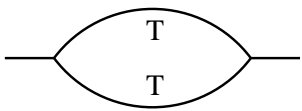}}
\vskip 5mm
\caption{Leading contribution to the self energy $\Sigma$. The internal
 propagators are given by the transverse velocity autocorrelation function.}
 \vskip 0mm
\label{fig:2}
\end{figure}

We first define the velocity autocorrelation function, 
\be
C_{\alpha\beta}({\bf k},t) = \langle u_{\alpha}({\bf k},t)\,
    u_{\beta}(-{\bf k},0)\rangle\quad,  
\label{eq:2.5}
\ee
with ${\bf k}$ the spatial Fourier transform variable. An equation for $C$
can be obtained by Fourier transforming Eq.\ (\ref{eq:2.4}), 
multiplying by $u_{\beta }(-{\bf k},0)$, and averaging over the noise while
keeping in mind
that the noise is uncorrelated with the initial fluid velocity. In the case
of an incompressible fluid we need to consider only
the transverse-velocity correlation function, $C_{\perp}$.
This is easily done by multiplying with unit vectors,
${\hat k}_{\perp}^{(i)}$ $(i=1,\ldots,d-1)$, that are perpendicular to
${\bf k}$, which eliminates the pressure term. We obtain
\bea
(\partial_t + \nu{\bf k}^2)\,C_{\perp}({\bf k},t)&=&
   -i\gamma k_{\mu}{\hat k}_{\perp\alpha}^{(i)} {\hat k}_{\perp\beta}^{(i)}
   \sum_{\bf q}\langle u_{\mu}({\bf k}-{\bf q},t)\,
\nonumber\\
&&\times u_{\alpha}({\bf q},t)\,u_{\beta}(-{\bf k},0)\rangle\quad,  
\label{eq:2.6}
\eea
where we have used the incompressibility condition, Eq.\ (\ref{eq:2.3}),
to write all gradients as external ones.
To zeroth order in $\gamma$ we find, with the help of the f-sum 
rule\cite{Forster,TCF_footnote} $C_{\perp}({\bf k},t=0) = k_{\rm B} T/\rho$,
\be
C_{\perp}({\bf k},t) = (k_{\rm B} T/\rho)\,e^{-\nu {\bf k}^2 t}+O(\gamma)\quad.
\label{eq:2.7}
\ee
This is the standard result obtained from linearized hydrodynamics, which
predicts exponential decay for ${\bf k}\neq 0$. 
To calculate corrections due to the
nonlinearity, we need an equation for the
three-point correlation function in Eq.\ (\ref{eq:2.6}). The simplest way
to do this is to use the time 
translational invariance property of this correlation function to put the 
time dependence in the last velocity, and then use Eq.\ (\ref{eq:2.4}) 
again.\cite{TCF_footnote} The result is an equation for the
three-point function in terms of a four-point one that is analogous to
Eq.\ (\ref{eq:2.7}). To solve this equation, we note
that, due to the velocity being odd under time reversal, the equal-time
three-point correlation function vanishes. By means of a Laplace transform,
one can therefore express the three-point function as a product (in frequency
space) of the zeroth order result for $C_{\perp}$, Eq.\ (\ref{eq:2.7}),
and the four-point function. To leading (i.e., zeroth) order in $\gamma$ the
latter factorizes into products of velocity autocorrelation functions.
Upon transforming back into time space, and
to quadratic order in the coupling constant $\gamma$, we thus 
obtain\cite{ZM_footnote}
\bml
\label{eqs:2.8}
\be
\left(\partial_t + \nu{\bf k}^2\right)\,C_{\perp}({\bf k},t) 
   + \int_0^t d\tau\ \Sigma ({\bf k},t-\tau)\,C_{\perp}({\bf k},\tau) = 0\quad,
\label{eq:2.8a}
\ee
with
\bea
\Sigma ({\bf k},t)&=&\gamma^2\,\frac{\rho}{k_{\rm B}T}\,k_{\mu}k_{\nu}
   {\hat k}^{(i)}_{\perp,\alpha}{\hat k}^{(i)}_{\perp,\beta}\sum_{\bf p}
   \Bigl[C_{\alpha\beta}({\bf p},t)\,
\nonumber\\
&&\times C_{\mu\nu}({\bf k}-{\bf p},t)
     + C_{\alpha\nu}({\bf p},t)\,C_{\mu\beta}({\bf k}-{\bf p},t)\Bigr]
\nonumber\\
&&\hskip 120pt     + O(\gamma^3)\quad.
\label{eq:2.8b}
\eea
\eml%
The self-energy $\Sigma$ is proportional to ${\bf k}^2$, and thus provides
a renormalization of the bare viscosity $\nu$. For later reference we
mention that diagrammatically the above contribution to $\Sigma$ can
be represented by the one-loop diagram shown in Fig.\ \ref{fig:2}.
Equation (\ref{eq:2.8b}), and its representation by Fig.\ \ref{fig:2},
illustrate the meaning of the term `mode-coupling' as explained in the
Introduction.

The source of the LTTs is now evident. In our model of an incompressible
fluid, only the transverse component of $C_{\alpha\beta}$ is nonzero,
\be
C_{\alpha\beta}({\bf p},t) = \left(\delta_{\alpha\beta} - {\hat p}_{\alpha}
   {\hat p}_{\beta}\right)\,C_{\perp}({\bf p},t)\quad.
\label{eq:2.9}
\ee
Putting $\gamma=1$, and defining 
$\delta\nu(t) = \lim_{{\bf k}\rightarrow 0}\,\Sigma({\bf k},t)/{\bf k}^2$,
one obtains for asymptotically long times
\be
\delta\nu(t) = \frac{k_{B}T}{\rho }\left( \frac{d^{2}-2}{d(d+2)}\right) 
\frac{1}{(8\pi\nu t)^{d/2}}\quad.  
\label{eq:2.10}
\ee
This is the well-known contribution of the transverse-velocity modes
to the LLT of the viscosity.\cite{EHvL_2} In a compressible fluid,
a similar process coupling two longitudinal modes also contributes to
the leading LTT. The other transport coefficients, e.g., $\lambda$ and
$\zeta$, also show LTTs proportional to $1/t^{d/2}$, and all of them
have also less leading LTTs proportional to $1/t^{(d+1)/2}$ or 
weaker.\cite{Dorfman,EHvL_2} 

The correction to the kinematic viscosity is given by the time
integral over $\delta\nu(t)$, and for the frequency-dependent kinematic
viscosity the Fourier transform of $\delta\nu(t)$ implies a nonanalyticity
at zero frequency. More generally,
for the frequency or wavenumber-dependent transport coefficients, the 
LTTs imply the asymptotic forms
\bml
\label{eqs:2.11}
\bea
\mu (\omega)/\mu (0)&=&1 - c^{\mu}_{d}\,\omega^{(d-2)/2} + \ldots\quad,
\label{eq:2.11a}\\
\mu ({\bf k})/\mu (0)&=&1 - b^{\mu}_{d}\vert{\bf k}\vert^{d-2} + \ldots\quad,
\label{eq:2.11b}
\eea
\eml%
where $\mu$ represents any of the transport coefficients, 
$\nu$, $\zeta$, $\lambda$, or 
$D$, in a fluid. The prefactors $c^{\mu}_d$ and $b^{\mu}_d$ are positive.

Of particular interest for the critical dynamics near the liquid-gas critical
point is a contribution to $\delta\lambda$ from the coupling of the
transverse-velocity fluctuations to the entropy fluctuations. Before 
performing the wavenumber integral, this particular contribution, 
which we denote by $\delta\lambda_{\perp s}$, 
is\cite{Kawasaki_in_DG,HohenbergHalperin} 
\be
\delta\lambda_{\perp s}({\bf k},t) = \frac{1}{\rho^2}\sum_{\bf p}\chi({\bf p})
   \sum_{i}({\hat{\bf k}}\cdot{\hat{\bf p}}_{\perp}^{(i)})^2\,
      e^{-(\nu{\bf p}_{-}^2 + D_T{\bf p}_{+}^2)\,t}\quad.
\label{eq:2.12}
\ee
Here $D_T = \lambda/\rho\,c_p$ is the thermal diffusivity in terms of
$c_p$, the specific heat at constant pressure. $\chi$ is the order
parameter susceptibility for the phase transition, where we have anticipated 
that near the critical point we will need the momentum-dependent $\chi$,
and ${\bf p}_{\pm} = {\bf p}\pm{\bf k}/2$. Setting ${\bf k}=0$ and carrying
out the momentum integral leads to a $t^{-d/2}$ LTT. The correction to
the thermal conductivity is obtained by integrating $\delta\lambda (t)$
over all times, as in the case of the kinematic viscosity.

These results imply that for $d\leq 2$, conventional hydrodynamics does not
exist. Indeed, it is now known that for these dimensions the hydrodynamic
equations are nonlocal in space and time. For a discussion of this topic we 
refer the reader elsewhere.\cite{2d_hydrodynamics} 

By examining Eq.\ (\ref{eq:2.8b}) or (\ref{eq:2.12}) one easily identifies a
mechanism by which the LTT effects can become even stronger. Consider
a system with long-range static correlations, for instance due to 
Goldstone modes, or due to the vicinity of a continuous phase transition. 
In either case, some susceptibilities, 
e.g. the $\chi$ in Eq.\ (\ref{eq:2.12}), 
become long-ranged, amplifying the LTT effect. In the next section we 
discuss the realization of this scenario in the vicinity of a phase 
transition. We note that in certain liquid crystal phases some susceptibilities
behave as $1/{\bf k}^2$, amplifying the LTT effect and
causing a breakdown of local hydrodynamics for all dimensions 
$d\leq 4$.\cite{MRT}

We make one last point concerning the LTTs. So far we have
stressed the leading tails that decay like $t^{-d/2}$, but their are numerous
subleading LTTs as well. Most of them are uninteresting, but 
one becomes important near the critical point, via the mechanism discussed
in the last paragraph. According to Eq.\ (\ref{eq:2.12}), 
a central quantity for determining the critical contribution to $\lambda$ 
is the shear viscosity $\eta$ (which enters $\nu$). It turns out that
the contribution to $\eta$ that is dominant near the critical point
is a subleading LTT away from criticality. It involves a coupling of two heat 
or entropy modes and is given by\cite{Kawasaki_in_DG}
\bea
\delta\eta ({\bf k},t)&=&\frac{A}{{\bf k}^2}\sum_{\bf p}\chi ({\bf p})\,
   \chi ({\bf k}-{\bf p})\left(\frac{1}{\chi({\bf p})}-\frac{1}{\chi({\bf k}
      -{\bf p})}\right)^2
\nonumber\\
&&\times ({\hat{\bf k}}_{\perp}\cdot{\bf p})^2\,e^{-D_T[{\bf p}^2 
       + ({\bf k}-{\bf p})^2)]t}\quad,
\label{eq:2.13}
\eea
with $A$ a constant. Away from the critical point, this LTT decays as 
$t^{-(d/2+2)}$, so in fact it is a next-to-next leading LTT. Nevertheless,
it is the dominant mode-coupling contribution to $\eta$ near the critical
point because of the two factors of $\chi $ in the numerator of 
Eq.\ (\ref{eq:2.13}). We discuss this enhancement of LTTs by critical
susceptibilities further in the next subsection.

\subsection{Critical dynamics}
\label{subsec:II.C}

Near continuous phase transitions, fluctuations grow and ultimately diverge
at the critical point. For the liquid-gas critical point the order parameter
is the difference between the density and the critical density,
and the divergent fluctuations are the density
fluctuations as described by the density susceptibility. 
The susceptibility $\chi$ in Eq.\ (\ref{eq:2.12})
is proportional to this divergent susceptibility, which in the
Ornstein-Zernike approximation behaves as 
\be
\chi({\bf p})\propto\frac{1}{{\bf p}^2 + \vert r\vert}\quad, 
\label{eq:2.14}
\ee
with ${\bf p}$ measured in suitable units, and $r$ the 
dimensionless distance from the critical point. Carrying out the
time integral, the leading singular contribution to the static,
wavenumber-dependent thermal conductivity is\cite{HohenbergHalperin}
\be
\delta\lambda ({\bf k}) = \frac{1}{\rho^2}\sum_{\bf p}\chi ({\bf p})\,
   \frac{\sum_{i}({\hat{\bf k}}\cdot{\hat{\bf p}}_{\perp}^{(i)})^2}
      {\nu{\bf p}_{-}^2 + D_T{\bf p}_{+}^2}\quad. 
\label{eq:2.15}
\ee
Using Eq.\ (\ref{eq:2.14}) in this equation we see that the homogeneous thermal
conductivity is infinite at the critical point for all $d\leq 4$, diverging
as $\vert r\vert^{-(4-d)/2}$. This is result of the amplification of
the LTT by the critical fluctuations. 

By the same mechanism, we see that
Eq.\ (\ref{eq:2.14}) leads to a logarithmically singular contribution to
$\delta\eta$, if we take into account that Eq.\ (\ref{eq:2.15}) implies
that at the critical point, $D_T({\bf k}) \sim \vert{\bf k}\vert^{d-2}$.

\begin{figure}[htb]
\epsfxsize=70mm
\centerline{\epsffile{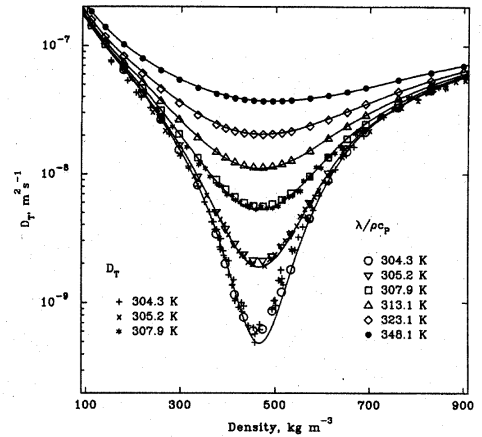}}
\vskip 5mm
\caption{The thermal diffusivity $D_T = \lambda/\rho\,c_p$ of carbon dioxide
 in the critical region as a function of density at various temeratures
 ($T_{\rm c} = 304.12\,{\rm K}$). The symbols indicate experimental data for
 $D_T$ measured directly, and for $\lambda/\rho\,c_p$ deduced from 
 thermal-conductivity data. The solid curves represent
 values calculated from the mode-coupling theory. From Ref.\
 \protect\onlinecite{LSO}.}
 \vskip 0mm
\label{fig:3}
\end{figure}

The above are one-loop calculations that use the Ornstein-Zernike
susceptibility. To go beyond these approximations one needs to (1) use the
correct scaling form for the susceptibility, and (2) use either a 
self-consistent mode-coupling theory,\cite{KadanoffSwift} or a
renormalization group approach\cite{ForsterNelsonStephen,HohenbergHalperin} 
to improve on the one-loop approximation. The result\cite{HohenbergHalperin} 
is that the thermal conductivity diverges like
\bml
\label{eqs:2.16}
\be
\lambda\propto\vert r\vert^{-0.57}\quad,  
\label{eq:2.16a}
\ee
and that the thermal diffusivity vanishes like
\be
D_T\propto\vert r\vert^{0.67}\quad.
\label{eq:2.16b}
\ee
\eml%
This is in good agreement with experimental results, as shown in 
Fig.\ \ref{fig:3}, which compares experimental and theoretical 
results for the thermal diffusivity of carbon dioxide in the 
critical region.

\subsection{Nonequilibrium effects}
\label{subsec:II.D}

We now turn to a fluid in a nonequilibrium steady state. It has been known
for some time that these systems in general exhibit GSI in spatial
correlations, and that the
order parameters for phase transitions in such systems couple to these
long-range correlations. The first study of a phase transition of this
type considered phase separation in a binary 
liquid under shear.\cite{Imeada_etal}
The spatial correlations responsible for the GSI are closely related
to those that lead to the LTTs in equilibrium time correlation functions.
We will consider a fluid in a steady, spatially uniform temperature
gradient $\nabla T$, but far away from any convective instability. 
Further, we use a number of approximations
that enable us to focus on the most interesting effects of such a temperature
gradient. For a justification of this procedure, as well as the underlying
details, we refer to Ref.\ \onlinecite{KCD}.

Focusing on the coupling between fluctuations of the transverse fluid 
velocity ${\bf u}_{\perp}$ and temperature fluctuations $\delta T$, 
and neglecting the nonlinearity in Eq.\ (\ref{eq:2.1b}), 
Eqs.\ (\ref{eqs:2.1}) can be written
\bml
\label{eqs:2.17} 
\bea
\partial_t u_{\perp,\alpha}&=&\nu\nabla^2 u_{\perp,\alpha} + \frac{1}{\rho_0}\,
   (\nabla_{\beta}P_{\alpha\beta})_{\perp}\quad,  
\label{eq:2.17a}\\
\partial_t\,\delta T + ({\bf u}\cdot\nabla) T&=& D_T\nabla^2\delta T
   + \frac{1}{\rho_0 T_0 C_p}\,(\nabla\cdot{\bf q})\ ,  
\label{eq:2.17b}
\eea
\eml%
where $\rho_0$ and $T_0$ indicate average values. These bilinear equations 
can be solved by means of Fourier and Laplace transformations.
Focusing on static or equal-time correlations, one finds, 
for example,\cite{KCD}
\bml
\label{eqs:2.18}
\bea
\langle\vert\delta \rho({\bf k})\vert^2\rangle&=&\rho k_{\rm B} T\left(
   \frac{\partial\rho}{\partial p}\right)_T 
   + \frac{\rho k_{\rm B} T(\alpha_T{\hat{\bf k}}_{\perp}\cdot\nabla T)^2}
      {D_T (\nu + D_T){\bf k}^4}\quad,
\nonumber\\ 
\label{eq:2.18a}\\
&&\hskip -54pt \langle{(\hat {\bf k}}_{\perp}\cdot{\bf g}({\bf k}))\,
   \delta\rho(-{\bf k})
    \rangle = \rho k_{\rm B}T\alpha_T\,\frac{{\hat{\bf k}}_{\perp}\cdot\nabla T}
       {(\nu + D_T){\bf k}^2}\quad,  
\label{eq:2.18b}
\eea
with, 
\be
\alpha_T = \frac{-1}{\rho}\,\left(\frac{\partial\rho}
               {\partial T}\right)_p\quad, 
\label{eq:2.18c}
\ee
\eml%
the thermal expansion coefficient.

There are several remarkable aspects of these results. First, 
Eq.\ (\ref{eq:2.18a}) for the density correlations implies that the 
first term, which also exists in equilibrium, is delta-function correlated 
in real space, while the second term decays like 
${\rm const.} - \vert{\bf x}\vert$ in
three-dimensions, with $\vert{\bf x}\vert$ the distance in real 
space.\cite{SchmitzCohen,Ortiz_etal} Equation 
(\ref{eq:2.18b}) shows that the transverse momentum-density correlation
functions decays as $1/\vert{\bf x}\vert$ 
in three-dimensions. Both of these results
show that spatial correlations in a NESS exhibit GSI. Second, the right-hand
side of Eq.\ (\ref{eq:2.18b}) is essentially the integrand in 
Eq.\ ({\ref{eq:2.15}), which shows the
close connection between the LTTs, singularities in transport coefficients
near continuous phase transitions, and the GSI of spatial correlations in a
NESS.

The nonequilibrium fluctuations arise from a coupling between the temperature
fluctuations perpendicular to the temperature gradient and the
transverse-momentum (viscous) fluctuations parallel to the temperature
gradient.\cite{KCD,SchmitzCohen,LawSengers} The amplitudes of these
fluctuations can be measured by small-angle light-scattering experiments.
Specifically, the dynamical structure factor
$S_{\rho\rho}({\bf k},t) = \langle\delta\rho({\bf k},t)\,\rho(-{\bf k},0)
\rangle$, 
which is proportional to the scattering cross section, in a
NESS has the form\cite{DKS}
\bml
\label{eqs:2.19}
\be
S_{\rho\rho}({\bf k},t) = S_0\left[(1+A_T)\,e^{-D_T{\bf k}^2 t}
 - A_{\rm v}\,e^{-\nu{\bf k}^2 t}\right]\ ,
\label{eq:2.19a}
\ee
where $S_0$ is the structure factor in equilibrium, and
\bea
A_T = \frac{c_p\nu/D_T}{T(\nu^2-D_T^2)}\,\frac{({\hat{\bf k}}_{\perp}\cdot
       \nabla T)^2}{{\bf k}^4}\quad,
\label{eq:2.19b}\\
A_{\bf v} = \frac{c_p}{T(\nu^2-D_T^2)}\,\frac{({\hat{\bf k}}_{\perp}\cdot
       \nabla T)^2}{{\bf k}^4}\quad.
\label{eq:2.19c}
\eea
\eml%
For $t=0$ one
recovers the equal-time density correlation function, Eq.\ (\ref{eq:2.18a}).
The amplitudes $A_T$ and $A_{\rm v}$ are proportional to
$(\nabla T)^2/{\bf k}^4$, which has been verified by experiments,
see Fig.\ \ref{fig:4}.
The agreement, with no adjustable parameters, is excellent. Notice that the 
amplitude of the temperature fluctuations is enhanced by a factor of a hundred
compared to the scattering by an equilibrium fluid. 
\begin{figure}[htb]
\epsfxsize=60mm
\centerline{\epsffile{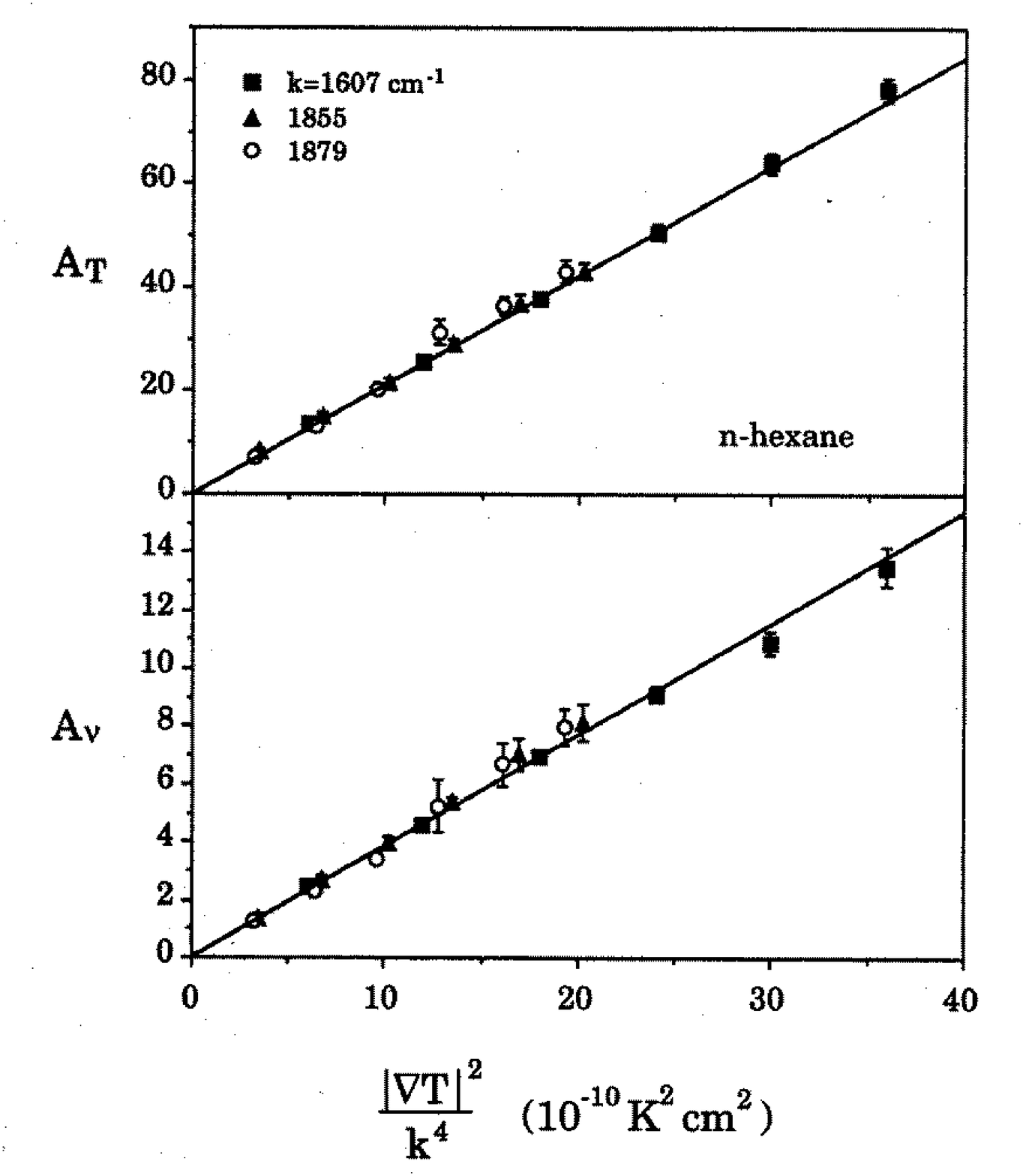}}
\vskip 5mm
\caption{Amplitudes $A_T$ and $A_{\rm v}$ of the nonequilibrium temperature
 and transverse-momentum (viscous) fluctuations in liquid hexane at 
 $25^o\,{\rm C}$ as a function of $(\nabla T)^2/{\bf k}^4$. The symbols
 indicate experimental data. The solid lines represent the values predicted
 by Eqs.\ (\ref{eqs:2.19}). From Ref.\ \protect\onlinecite{LSGS}.}
 \vskip 0mm
\label{fig:4}
\end{figure}
\begin{figure}[htb]
\epsfxsize=60mm
\centerline{\epsffile{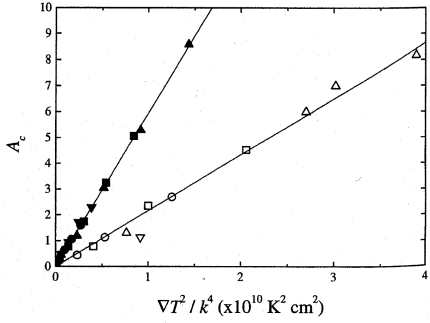}}
\vskip 5mm
\caption{Amplitude $A_c$ of nonequilibrium concentration fluctuations in
 polystyrene and toluene solutions at $25^o\,{\rm C}$ as a function of
 $(\nabla T)^2/{\bf k}^4$. The symbols indicate experimental data at
 polymer mass fractions $w=4.00\%$ (solid symbols) and $w=0.50\%$ (open
 symbols), respectively. Different symbol shapes correspond to different
 values of the wave number $k$. The solid lines are linear fits to the
 experimental data in agreement with the theoretical 
 predictions.\protect\cite{LZSGO}}
 \vskip 0mm
\label{fig:5}
\end{figure}

In a liquid mixture or a polymer solution, a temperature gradient induces
a concentration gradient due to the Soret effect. Nonequilibrium
concentration fluctuations then arise from a coupling between the
concentration fluctuations and the transverse-momentum
fluctuations.\cite{LawNieuwoudt,SengersOrtiz} Just as in the case of 
the nonequilibrium temperature and viscous fluctuations, the
amplitude of the nonequilibrium fluctuations will also be proportional
to $(\nabla T)^2/{\bf k}^4$. The presence of such long-range
nonequilibrium concentration fluctuations has also been confirmed
experimentally, see Fig.\ \ref{fig:5}.

One final point concerns the question what these generic long-range spatial 
correlations would do to critical behavior. This is a natural question, as
usually long-range interactions, or long-range correlations, fundamentally 
modify important aspects of any phase transition. This questions has been 
addressed elsewhere in the context of phase transition in nonequilibrium, 
or driven, classical systems.\cite{NESS} In the remainder of this paper we 
study the analogous question in the context of quantum, or zero-temperature, 
phase transitions in equilibrium systems.

\section{Generic scale invariance and phase transitions in quantum systems}
\label{sec:III}

\subsection{Field theory}
\label{subsec:III.A}

We now turn to a quantum fluid, specifically, to systems of interacting
electrons in solids. Since we will again be interested in effects at
long wavelengths and long times, the ionic lattice and effects resulting
from it will be irrelevant for our purposes, and we therefore adopt a
simple `jellium' model of free electrons interacting via the Coulomb
interaction  in the presence of a charge neutralizing homogeneous background.
There are various ways to theoretically deal with this many-body problem,
but the most powerful one for our purposes is a field-theoretic
formulation of quantum statistical mechanics. This method\cite{NegeleOrland}
starts by writing the partition function in terms of a functional integral,
\be
Z = \int_{\psi(0) = -\psi(1/k_{\rm B}T)} D[{\bar\psi},\psi]\ 
     e^{S[{\bar\psi},\psi]} \quad.
\label{eq:3.1}
\ee
Here $D$ is a functional integration measure with respect to fermionic, i.e.,
Grassmann valued, fields ${\bar\psi}$ and $\psi$ that are defined on a
real-space/imaginary-time manifold with the time sector spanning the
interval $[0,1/k_{\rm B}T]$. The action $S$ for our model system reads
\be
S = \int dx\ \sum_{\sigma}{\bar\psi}_{\sigma}(x)\,\left[-\partial_{\tau} 
     + \nabla^2/2m + \mu\right]\,\psi_{\sigma}(x) + S_{\rm int}\ .
\label{eq:3.2}
\ee
Here $x\equiv ({\bf x},\tau)$ with ${\bf x}$ denoting position, $\tau$
imaginary time, and $\sigma$ is the spin index. 
$\int dx \equiv \int d{\bf x} \int_0^{1/k_{\rm B}T}d\tau$, $m$ is
the electron mass, and $\mu$ is the chemical potential. At $T=0$,
$\mu = k_{\rm F}^2/2m$ with $k_{\rm F}$ 
the Fermi wavenumber. Here and in what follows
we use units such that $\hbar=1$. At the most basic level of the theory, 
$S_{\rm int}$ represents the Coulomb potential,
but it often is advantageous to integrate out certain degrees of freedom
to arrive at effective, short-ranged interactions.\cite{us_fermionsII}
For our purposes we do not need to specify the precise form of 
$S_{\rm int}$, it will suffice to postulate that the ground state of the
interacting system is a Fermi liquid. 

It is useful to go to a Fourier representation with wavevectors ${\bf k}$
and fermionic Matsubara frequencies $\omega_n = 2\pi T(n+1/2)$,
\bml
\label{eqs:3.3}
\bea
\psi_{n,\sigma}({\bf k})&=&\sqrt{T/V}\int dx\ 
   e^{-i({\bf kx} - \omega_n\tau)}\,\psi_{\sigma}(x)\quad,
\label{eq:3.3a}\\
{\bar\psi}_{n,\sigma}({\bf k})&=&\sqrt{T/V}\int dx\ 
   e^{i({\bf kx} - \omega_n\tau)}\,{\bar\psi}_{\sigma}(x)\quad.
\label{eq:3.3b}
\eea
\eml%
From Eq.\ (\ref{eq:3.2}) it follows that, in a free-electron system, 
single-particle excitations about the Fermi surface are soft 
with a linear dispersion,
\be
\langle{\bar\psi}_{n,\sigma}({\bf k})\,\psi_{m,\sigma'}({\bf p})
   \rangle_{q\rightarrow 0\atop \omega_n\rightarrow 0} =
    \delta_{{\bf k},{\bf p}}\,\delta_{nm}\,\delta_{\sigma\sigma'}\,
        \frac{Z}{i\omega_n - v_{\rm F}q}\quad,
\label{eq:3.4}
\ee
with $Z=1$, $q = \vert{\bf k}\vert - k_{\rm F}$, and
$v_{\rm F} = k_{\rm F}/m$ the Fermi velocity. Here
$\langle\ldots\rangle = \int D[{\bar\psi},\psi]\,\ldots\,e^S/Z$ comprises
both a quantum mechanical and a thermodynamic average.
Fermi liquid theory shows that Eq.\ (\ref{eq:3.4}) remains
valid in the presence of $S_{\rm int}$, only the values of $Z$ and 
$v_{\rm F}$ are changed compared to a free electron 
gas.\cite{FL} Similarly, there are soft
two-particle excitations. For instance, the susceptibility
\bea
\sum_{{\bf k},{\bf p}}\sum_{n',m'}& &
   \left\langle\left({\bar\psi}_{n,\sigma}({\bf k}+{\bf q})
   \psi_{m,\sigma}({\bf k})\right)\,\right.
\nonumber\\
&&\hskip 30pt \times\left.
   \left({\bar\psi}_{n',\sigma'}({\bf p}-{\bf q})
   \psi_{m',\sigma'}({\bf p})\right)\right\rangle
\label{eq:3.5}
\eea
diverges for ${\bf q}=0$, $nm<0$, and 
$\Omega_{n-m}\equiv\omega_n-\omega_m\rightarrow 0$
like $1/\vert\Omega_{n-m}\vert$. At nonzero external wavenumbers
$\vert{\bf q}\vert$, the wavenumber 
scales like the frequency, so the dispersion of the soft modes is again linear.

In many solid-state electronic systems, quenched disorder is present in the
form of impurities or lattice defects. This has important consequences for
the transport and thermodynamic properties of the system, and we model it
by means of a term in the action
\bml
\label{eqs:3.6}
\be
S_{\rm dis} = \int dx\ u({\bf x})\sum_{\sigma}{\bar\psi}_{\sigma}(x)\,
              \psi_{\sigma}(x)\quad.
\label{eq:3.6a}
\ee
Here $u({\bf x})$ is a static, random potential governed by a distribution
$P[u]$. For simplicity, we take $u({\bf x})$ to be delta-correlated and
Gaussian distributed with the second moment of $P$ given by
\be
\{u({\bf x})\,u({\bf y})\}_{\rm dis} = \frac{1}{\pi N_{\rm F}\tau_{\rm rel}}
  \ \delta({\bf x}-{\bf y})\quad,
\label{eq:3.6b}
\ee
\eml%
Here $\{\ldots\}_{\rm dis}$ denotes the disorder average, $\tau_{\rm rel}$
is the elastic relaxation time, and $N_{\rm F}$ is the density of states
at the Fermi surface. In the presence of $S_{\rm dis}$, the
single-particle excitations, Eq.\ (\ref{eq:3.4}), are obviously massive with
a mass proportional to $1/\tau_{\rm rel}$. However, two-particle excitations 
are still soft. In contrast to the clean case, their dispersion relation
is diffusive. For instance, the disorder averaged analog of the
two-particle propagator given in Eq.\ (\ref{eq:3.5}) has the form\cite{us_R}
\bea
\sum_{{\bf k},{\bf p}}\sum_{n',m'}&\Bigl\{&
   \bigl\langle\left({\bar\psi}_{n,\sigma}({\bf k}+{\bf q})
   \psi_{m,\sigma}({\bf k})\right)\,
\nonumber\\
     &&\hskip -30pt \times\left({\bar\psi}_{n',\sigma'}({\bf p}-{\bf q})
   \psi_{m',\sigma'}({\bf p})\right)\bigr\rangle\Bigr\}_{\rm dis}
   \propto \frac{1}{\vert\Omega_{n-m}\vert + D{\bf q}^2}\quad,
\nonumber\\
\label{eq:3.7}
\eea
where the diffusion coefficient $D$ is proportional to $\tau_{\rm rel}$.
In both the clean and the disordered cases, the two-particle excitations
have the same form for different combinations of the spin indices.

Notice that the softness of the two-particle excitations is not related
to any conservation law, i.e., the combinations of fermion fields do not
correspond to a density or, in the clean case, a current. Rather, these
excitations are the Goldstone modes resulting from the spontaneous
breaking of the symmetry between positive and negative Matsubara
frequencies. The order parameter that belongs to this symmetry is the
quantity
\bea
Q&=&\lim_{\omega_n\rightarrow 0+}\left\langle{\bar\psi}_{n,\sigma}({\bf x})\,
   \psi_{n,\sigma}({\bf x})\right\rangle
\nonumber\\
&&\hskip 50pt -
\lim_{\omega_n\rightarrow 0-}\left\langle{\bar\psi}_{n,\sigma}({\bf x})\,
   \psi_{n,\sigma}({\bf x})\right\rangle \quad,
\label{eq:3.8}
\eea
which is the single-particle spectral function, or the difference between
the retarded and advanced Green functions. Since the causal Green function
has a cut on the real axis, $Q$ is nonzero as long as the density of states
at the Fermi surface is nonzero. This connection between the single-particle
spectral function and the soft mode spectrum was discovered by Wegner for
the disordered case,\cite{Wegner} and has been elaborated on in 
Refs.\ \onlinecite{WegnerSchaefer,us_fermions,us_fermionsII}.

\subsection{Long-time tails and spatial GSI in equilibrium}
\label{subsec:III.B}

\subsubsection{Long-time tails, a.k.a. weak-localization effects}
\label{subsubsec:III.B.1}

At zero temperature, a clean electron system has an infinite diffusivity
or conductivity, so we start our discussion with the disordered case. One
expects the mode-coupling argument given in Sec.\ \ref{sec:II} to still
apply, with the transverse-velocity modes of Eqs.\ (\ref{eqs:2.8}) replaced
by the diffusive modes of Eq.\ (\ref{eq:3.7}). Since the dispersion relation
is the same in both the classical fluid case and the disordered electron
case, we thus expect a LTT in the real part of the conductivity that takes
the same functional form as Eq.\ (\ref{eq:2.11a}), viz.
\be
\sigma(\omega)/\sigma_0 = 1 - c^{\sigma}_d\,\omega^{(d-2)/2}\quad.
\label{eq:3.9}
\ee
This expectation is indeed borne out by an explicit calculation for both
noninteracting\cite{GLK} and interacting\cite{AAL} electrons. It turns out,
however, that the coefficient $c^{\sigma}_d$ is {\em negative}, in contrast
to the classical case. This difference in the sign of the LTT is due to the
fact that the scatterers that lead to a finite conductivity in the quantum
case are static, while in the classical fluid they are the moving fluid
particles themselves. The strength of the LTT, on the other hand, is the
same in both cases due to the strong similarity of the respective soft
mode spectra.

In this context it is interesting to point out that another, and in some
sense closer, classical analog of the quenched disordered electron fluid
is the classical Lorentz gas, which consists of noninteracting classical
particles moving between static scatterers.\cite{Hauge} The diffusivity
in this case also shows a LTT with the same sign as in the quenched disordered
quantum case (and for the same reason), but the strength of the LTT is
weaker, viz. $\omega^{d/2}$ in frequency space or $1/t^{(d+2)/2}$ in time
space. The reason is that in the classical Lorentz gas there is no analog
of the spontaneously broken symmetry mentioned above, and the only soft mode
(which is diffusive) is due to particle number conservation.\cite{Ernst_et_al}
There are thus many fewer soft modes than in either a classical fluid or a
disordered electron system, which leads to a weaker mode-coupling effect and,
hence, to a weaker LTT. Indeed, at nonzero temperature the $\omega^{(d-2)/2}$
dependence in Eq.\ (\ref{eq:3.9}) gets transformed into a $T^{(d-2)/2}$
behavior, and in the limit $\omega\rightarrow 0$ at fixed $T>0$ one recovers
the result for the classical Lorentz gas. This example illustrates that for
the purpose of comparing LTT effects in different systems the crucial
criterion is how similar the soft-mode spectra are. In this respect the
disordered electron system, even without electron-electron interactions,
is closer to the classical fluid than to its classical limit, the classical
Lorentz gas. 

The fact that a nonzero temperature cuts off the LTT singularity
provides the most convenient way to experimentally observe the phenomenon
in electronic systems, since the dynamical conductivity is hard to measure.
Fig.\ \ref{fig:6} shows the $\sqrt{T}$ dependence of $\sigma(\omega=0,T)$
\begin{figure}[htb]
\epsfxsize=50mm
\centerline{\epsffile{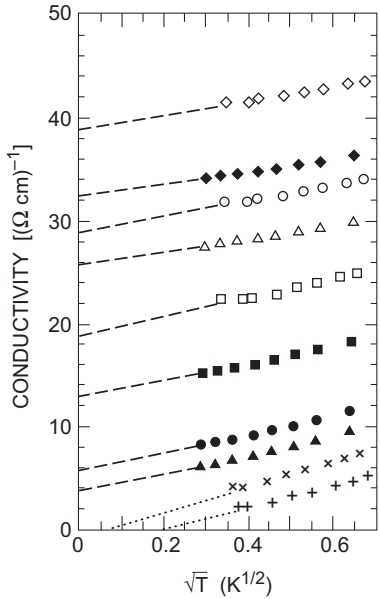}}
\vskip 5mm
\caption{Conductivity data from Ref.\ \protect\onlinecite{Dai_etal} showing
 the static conductivity of ten Si:B samples in a magnetic field plotted
 against $\protect\sqrt{T}$ at low temperatures. 
 From Ref.\ \protect\onlinecite{us_R}.}
 \vskip 0mm
\label{fig:6}
\end{figure}
in bulk Si:B. In $2$-$d$ systems, realized by thin metallic films, the
zero exponent in Eq.\ (\ref{eq:3.9}) translates into a logarithmic frequency
or temperature dependence within a large dynamical range, like in the
classical case, Eq.\ (\ref{eq:1.2b}). An experimental
example of this phenomenon is shown in Fig.\ \ref{fig:7}.
\begin{figure}[htb]
\epsfxsize=60mm
\centerline{\epsffile{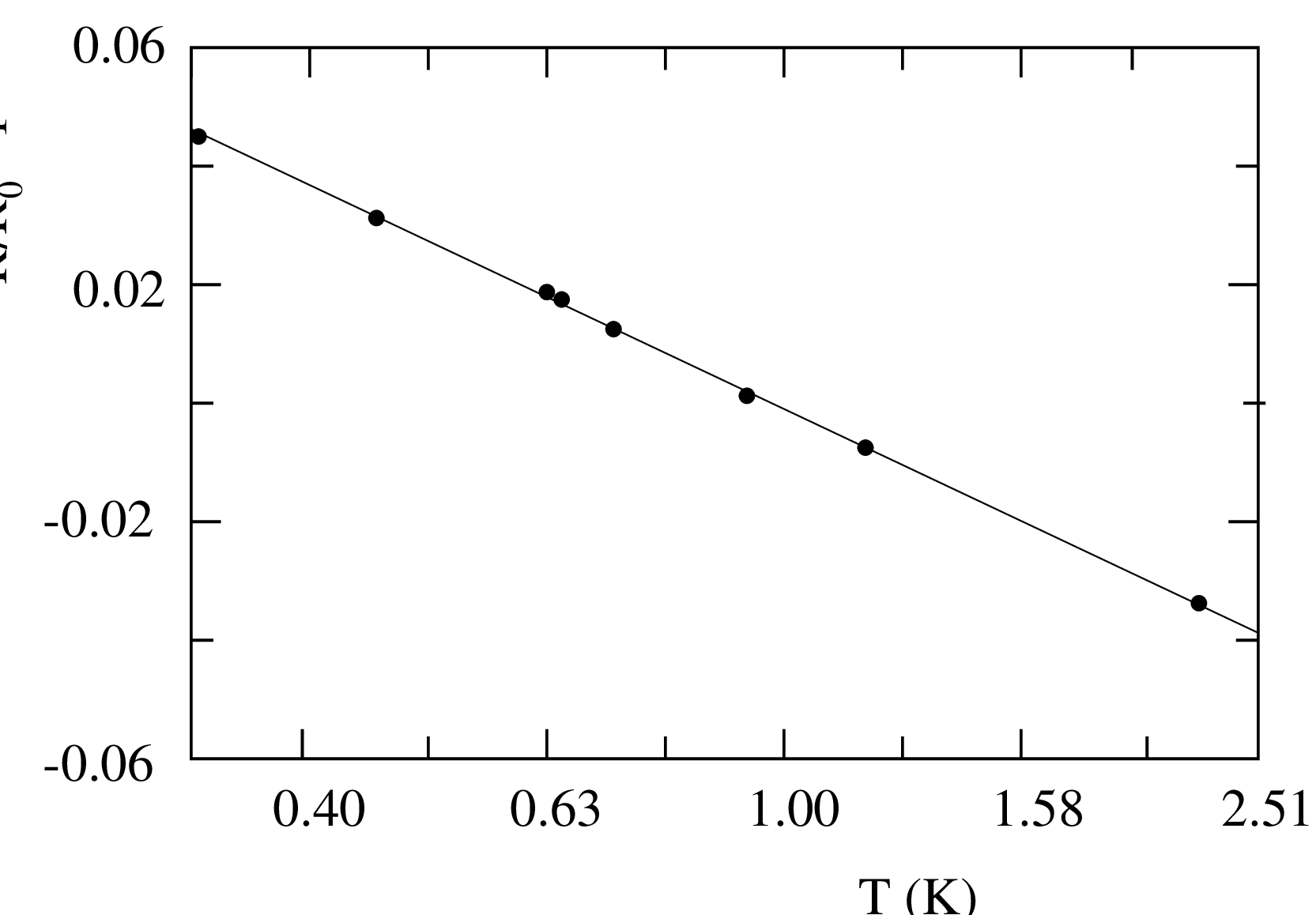}}
\vskip 5mm
\caption{Resistance data from Ref.\ \protect\onlinecite{DolanOsheroff},
 for a thin PdAu film. The resistance $R$, normalized to $R_0 = R(T=1{\rm K})$,
 is plotted versus $\log T$. From Ref.\ \protect\onlinecite{us_ernst}.}
 \vskip 0mm
\label{fig:7}
\end{figure}

These LTTs in disordered electron systems are known as `weak-localization
effects'. This is because they can be considered precursors of the
`strong localization' that occurs in systems with very strong disorder,
which undergo a phase transition from a metal to an insulator as a function
of the disorder.\cite{Anderson} In two-dimensions the weak-localization 
effects are strong enough to prevent the formation of a true metal 
altogether,\cite{gang_of_4} and as a function of increasing disorder 
the system undergoes a crossover from a weak
insulator to a strong one. Systems in $d\leq 2$ are always localized,
or insulating, irrespective of the strength of the disorder, at least
for noninteracting electrons.\cite{2d_MIT_footnote}

\subsubsection{Generic scale invariance in equilibrium}
\label{subsubsec:III.B.2}

In quantum statistical mechanics, as opposed to the classical theory, there
is an intrinsic coupling between statics and dynamics. This is manifested by
the basic fermionic fields ${\bar\psi}$ and $\psi$ being functions of both
space and (imaginary) time. Consequently, one expects effects similar to
the LTTs in time correlation functions to appear in static correlation
functions, even in equilibrium. As an example, let us consider the wavenumber 
dependent static spin susceptibility $\chi_{\rm s}$ 
in a disordered electron system.
Given that the wavenumber scales like the square root of the 
frequency, Eq.\ (\ref{eq:3.7}), it is natural to guess that $\chi_{\rm s}$ at
small wavenumbers has the form
\be
\chi_{\rm s}({\bf q})=c_0-c_{d-2}\vert{\bf q}\vert^{d-2} + O({\bf q}^2)\quad.
\label{eq:3.10}
\ee
The coefficients $c_0$ and $c_{d-2}$ are both expected to be positive.
This is because the quenched disorder slows the electrons down, which
leads to an increased effective electron-electron interaction. Consequently,
the disorder enhances $c_0$ compared to the value of the homogeneous
spin susceptibility in a clean system, and the susceptibility decreases
with increasing wavenumber. This is the same effect that leads to a
positive coefficient $c_d^{\sigma}$ in Eq.\ (\ref{eq:3.9}).
This result is confirmed by explicit calculations.\cite{us_fm_dirty} 
This nonanalytic wavenumber dependence, which for
dimensions $2<d<4$ is the dominant one, corresponds to a $1/r^{2(d-1)}$
decay in real space. We see that the coupling between statics and
dynamics produce GSI in quantum systems even in equilibrium. Note that
this effect is much weaker than the GSI seen in non-equilibrium classical
(see Sec.\ \ref{subsec:II.D}) or quantum (see Sec.\ \ref{subsec:III.D} 
below) systems.

Detailed calculations show that the origin of the GSI can be traced, as
the above argument suggests, to the same Goldstone modes that lead to the
LTTs and were discusssed in the previous subsection. The relevant
contribution to $\chi_{\rm s}$ can be schematically represented by
the integral\cite{us_fm_dirty}
\be
\int_{\vert{\bf q}\vert}^{\Lambda} dp\,p^{d-1}
   \int d\omega\ \frac{\omega}{(\omega + T + p^2)^3}\quad,
\label{eq:3.11}
\ee
with $\Lambda$ a microscopic wavenumber. Equation\ (\ref{eq:3.11})
demonstrates the coupling between statics and dynamics mentioned above.
Notice that for power-counting purposes the integrand is again a
product of two diffusive modes.
At $T=0$, the integral yields Eq.\ (\ref{eq:3.10}), and at
$T>0$ the temperature cuts off the leading singularity as discussed
in Sec.\ \ref{subsubsec:III.B.1} above. In contrast to the LTT of
the previous subsection, the above results hold for interacting electron
systems only. For noninteracting electrons, the absence of frequency
mixing prevents the coupling of the static spin density fluctuations 
to the soft modes, and $\chi_{\rm s}$ is analytic at zero wavenumber.

One might expect all static correlation functions to display such nonanalytic
wavenumber dependences, but this turns out not to be true. For instance,
the particle number density susceptibility does not have a leading 
singularity analogous to that in $\chi_{\rm s}$, and neither 
does any other spin-singlet particle-hole
susceptibility like, e.g., the number density current susceptibility. This
is because these observables couple less strongly to the soft modes than
the spin density. In Ref.\ \onlinecite{us_OP} general criteria have been
developed that allow to determine which susceptibilities show GSI
due to soft modes, and which do not. 
This reference also discusses limitations of the
equivalency between statics and dynamics in quantum statistical mechanics.
For instance, it is remarkable that the conductivity in a disordered
electron systems shows an $\omega^{(d-2)/2}$ LTT, see Eq.\ (\ref{eq:3.9}),
but the corresponding static current susceptibility does not show an analogous
$\vert{\bf q}\vert^{d-2}$ wavenumber dependence. This is because a finite
frequency breaks the symmetry in frequency space that we mentioned in
Sec.\ \ref{subsec:III.A}. As a result, dynamical current fluctuations couple
more strongly to the diffusive soft modes than static ones.

We now turn to the clean case. As we have seen, the only difference compared
to the disordered case is the dispersion relation of the soft modes, which
now is ballistic rather than diffusive, i.e., wavenumbers scale like 
frequencies, which leads to weaker LTT and GSI effects. Let us again consider
the static spin susceptibility. From Eq.\ (\ref{eq:3.11}) we expect that,
at zero temperature, there is a contribution to $\chi_{\rm s}$ that is of the
form
\bml
\label{eqs:3.12}
\be
\int_{\vert{\bf q}\vert}^{\Lambda} dp\,p^{d-1}
   \int d\omega\ \frac{\omega}{(\omega + p)^3}\quad.
\label{eq:3.12a}
\ee
wich leads to
\be
\chi_{\rm s}({\bf q}) 
   = c_0' + c_{d-1}\vert{\bf q}\vert^{d-1} + O({\bf q}^2)\quad,
\label{eq:3.12b}
\ee
\eml%
with positive coefficients $c_0'$ and $c_{d-1}$. Explicit calculations
confirm this result.\cite{us_chi_s,Millis} As in the case of the LTTs,
the signs of the nonanalyticities in the clean and disordered cases,
respectively, are different. A simple physical explanation in the present
case is that the nonanalyticity is produced by fluctuation effects that
weaken the tendency towards ferromagnetism. Consequently, they decrease
the value of $\chi_{\rm s}({\bf q}=0)$, and with increasing wavenumber
the susceptibility increases.

\subsection{Quantum phase transitions}
\label{subsec:III.C}

In Sec.\ \ref{subsec:II.C} we have seen how LTTs affect the critical behavior
at a classical phase transition. Here we discuss the analogous problem for
a quantum phase transition.\cite{Sachdev,Sondhi_etal,us_March}
An example is the ferromagnetic transition that is observed in, e.g.,
MnSi,\cite{Pfleiderer_etal} UGe$_2$,\cite{UGe_2} or ZrZn$_2$\cite{ZrZn_2} 
at $T=0$ as a function of hydrostatic pressure.
A crucial difference between the quantum and classical cases is that in
the latter, due to the lack of coupling between statics and dynamics,
the LTTs affect only the critical dynamics and the critical behavior of
quantities given by time correlation functions, like the thermal conductivity
in the example in Sec.\ \ref{subsec:II.C}. In the quantum case, however,
the LTT/GSI phenomena also influence the critical behavior of thermodynamic
quantities. The easiest way to see this is by deriving an order-parameter
or Landau-Ginzburg-Wilson (LGW) theory for the phase 
transition.\cite{WilsonKogut}

\subsubsection{Order-parameter field theory}
\label{subsubsec:III.C.1}

Let us consider a phase transition with an order-parameter field 
$n(x) = n({\bar\psi}(x),\psi(x))$ that is bilinear in the fermion fields. 
For example, in the ferromagnetic case, $n$ is a vector
field, viz. the spin density $n_i(x) = \sum_{\sigma,\sigma'}
{\bar\psi}_{\sigma}(x)(\sigma_i)_{\sigma,\sigma'}\psi_{\sigma'}(x)$,
with $\sigma_i$ ($i=x,y,z$) the Pauli matrices. In general, $n$ is a
rank-$m$ tensor field. In order for a phase transition to a phase with
a nonvanishing expectation value of $n$ to occur, the interaction term
$S_{\rm int}$ in the action must contain an interaction between the
order-parameter modes. This term, which we denote by $S_{\rm int}^{\rm OP}$,
reads schematically
\be
S_{\rm int}^{\rm OP} = \Gamma \int dx\ n^2(x)\quad,
\label{eq:3.13}
\ee
with $\Gamma$ a coupling constant. For simplicity, here and in what follows we 
suppress both the spin and tensor labels, and use an obvious symbolic
notation. The full action we write as
\be
S = S_0 + S_{\rm int}^{\rm OP}\quad,
\label{eq:3.14}
\ee
with the action $S_0$, which we will refer to as the `reference ensemble',
containing all contributions other than $S_{\rm int}^{\rm OP}$.

We now follow Hertz\cite{Hertz} in deriving an order-parameter field theory
for the phase transition under consideration. To this end, we first
decouple $S_{\rm int}^{\rm OP}$ by means of a Hubbard-Stratonovich
transformation.\cite{HS} Denoting the Hubbard-Stratonovich field by $M$, we
write the partition function
\bea
Z&=&\int D[{\bar\psi},\psi]\ e^{S[{\bar\psi},\psi]}
\nonumber\\
 &=&{\rm const.}\hskip -1pt\times\hskip -4pt\int\hskip -4pt
     D[M]\ e^{-\Gamma\int dx\,M^2(x)}\left\langle
     e^{- 2\Gamma\hskip -1pt \int\hskip -1pt dx M(x)\,n(x)}\right\rangle_{0},
\nonumber\\
 &\equiv&{\rm const.}\times\int D[M]\ e^{-\Phi[M]}\quad,
\label{eq:3.15}
\eea
where $\langle\ldots\rangle_0$ denotes an average with the reference ensemble
action $S_0$, and $\Phi[M]$ is the LGW functional. The latter reads explicitly
\be
\Phi[M] = \Gamma\int dx\ M^2(x) - \ln\left\langle e^{-2\Gamma\int dx\ 
          M(x)\,n(x)} \right\rangle_0\quad,
\label{eq:3.16}
\ee
and can be expanded in powers of $M$,
\bml
\label{eqs:3.17}
\bea
\Phi[M]&=&\frac{1}{2}\int dx_1\,dx_2\ M(x_1)\biggl[\frac{1}{\Gamma}\,
          \delta(x_1-x_2)
\nonumber\\
&&\hskip 50pt - \chi^{(2)}(x_1-x_2)\biggr]\,M(x_2)
\nonumber\\
&&\hskip 5pt +\frac{1}{3\,!}\int dx_1\,dx_2\,dx_3\ \chi^{(3)}(x_1,x_2,x_3)
\nonumber\\
&&\hskip 19pt \times M(x_1)\,M(x_2)\,M(x_3) + O(M^4)\quad,
\label{eq:3.17a}
\eea
where we have scaled $M$ with $1/\sqrt{2}\,\Gamma$. 
The coefficients $\chi^{(l)}$
in the Landau expansion, Eq.\ (\ref{eq:3.17a}), are connected $l$-point
correlation functions of $n(x)$ in the reference ensemble,
\be
\chi^{(l)}(x_1,\ldots,x_l) = \left\langle n(x_1)\cdots n(x_l)
                              \right\rangle_0^c\quad.
\label{eq:3.17b}
\ee
\eml

\subsubsection{Local versus nonlocal LGW theories}
\label{subsubsec:III.C.2}

A crucial question now arises concerning the behavior of the correlation
functions defined in Eq.\ (\ref{eq:3.17b}). Suppose their Fourier
transforms are finite in the limit of small wavenumbers and frequencies.
In lowest order in a gradient expansion one can then simply localize
the cubic and higher-order coefficients. Defining a Fourier transform
of the Hubbard-Stratonovich field by
\be
M(q) = \sqrt{T/V} \int dx\ e^{-i{\bf q}{\bf x} + i\omega_n\tau}\
       M(x)\quad.
\label{eq:3.18}
\ee
we obtain an LGW functional
\bea
\Phi[M]&=&\frac{1}{2}\sum_q M(q)\,\left[1/\Gamma - \chi^{(2)}(q)
          \right]\,M(-q)
\nonumber\\
&& + \frac{u_4}{4!}\frac{T}{V}\sum_{q_1,\ldots,q_4}\delta(q_1+q_2
          +q_3+q_4)\,M(q_1)\,M(q_2)
\nonumber\\
&&\hskip 50pt \times M(q_3)\,M(q_4) + \ldots
\label{eq:3.19}
\eea
Here the four-vector $q = ({\bf q},\omega_n)$ comprises wavevector
and Matsubara frequency, and we have assumed that $\chi^{(3)}$ vanishes
at zero frequency and zero wavenumber. Suppose $\chi^{(2)}$ is an 
analytic function of the wavenumber,
\be
\chi^{(2)}(q) = c_0 + c_2{\bf q}^2 + c_{\Omega}\vert\Omega_n\vert\,/\vert
                {\bf q}\vert^m + \ldots\quad,
\label{eq:3.20}
\ee
where the value of $m$ depends on the specific system. Then we obtain an
ordinary Landau theory with the Gaussian part of the LGW functional given by
\bea
\Phi^{(2)}[M]&=&\frac{1}{2}\sum_q M(q)\,\left[r + a_2\,{\bf q}^2
                + a_{\Omega}\vert\Omega_n\vert/\vert{\bf q}\vert^m\right]
\nonumber\\
             &&\hskip 50pt \times M(-q)\quad.
\label{eq:3.21}
\eea
Here $r = 1/\Gamma - \chi^{(2)}(q=0)$ is the bare distance from the
critical point at $T=0$. 
Equation (\ref{eq:3.21}) leads to mean-field 
values for 
the critical exponents $\nu$ and $\gamma$, and a dynamical critical
exponent $z=m+2$. Power counting shows that the non-Gaussian terms are
irrelevant in the renormalization-group sense of the word and, hence,
the exact critical behavior is also mean field-like. This is the
conclusion that was reached by Hertz,\cite{Hertz} namely, that generically
quantum phase transitions have mean-field critical behavior.

As we have seen in Sec.\ \ref{subsubsec:III.B.2}, however, this scenario
does not necessarily apply. Equation (\ref{eq:3.10}) provides an example
of a susceptibility that does not have the form of Eq.\ (\ref{eq:3.20})
due to GSI effects. Moreover, if the nonanalytic wavenumber dependence
of the susceptibility $\chi^{(2)}$ is caused by soft modes that are
made massive by an external field $H$ conjugate to the order parameter
(or by a nonvanishing average of the order parameter), then the
singular part of the static, field-dependent susceptibility will have the form
\be
\chi^{(2)}_{\rm sing}({\bf q},\Omega_n=0,H) = (\vert{\bf q}\vert^x + H)^y
    \quad,
\label{eq:3.22}
\ee
where $y>0$ and $x>0$ are exponents that determine the nature of the
nonanalyticity and the scaling of $H$ with the wavenumber, respectively.
This is true, for instance, in the case of the disordered itinerant
ferromagnet, where $x=2$ and $y=(d-2)/2$. Since the higher susceptibilities
$\chi^{(3)}$, $\chi^{(4)}$, etc., can be obtained from $\chi^{(2)}$ by
differentiating with respect to $H$, this implies that they diverge
in the limit of zero wavenumbers and frequencies. That is, the gradient
expansion that leads to the usual LGW theory, and hence LGW theory itself,
do not exist. Instead, one obtains a field theory where the coefficients
of the various powers of the field $M$ are singular functions of wavenumber
and frequency. In other words, the field theory is not local.\cite{us_fm_dirty} 
The above arguments make it clear that this is a direct consequence of the GSI 
in the equilibrium quantum system. 

Such nonlocal field theories are difficult to handle and not suitable for
explicit calculations. For the purpose of determining the critical behavior
one can try to circumvent this problem by checking whether the field theory
still allows for a Gaussian critical fixed point with respect to which the
higher-order terms are irrelevant by power counting. This was the
approach taken in Ref.\ \onlinecite{us_fm_dirty} for the disordered
ferromagnetic transition, and it relied strictly
on power counting by taking the divergence of, e.g., the coefficient $u_4$
in Eq.\ (\ref{eq:3.19}) into account by assigning it a suitable scale
dimension. It turns out that this procedure yields the correct power laws,
but misses logarithmic corrections to scaling that power counting is not
sensitive to. In the following subsection we use this example to illustrate
the interplay between GSI and quantum critical behavior.

\subsubsection{Example 1: The disordered quantum ferromagnetic transition}
\label{subsubsec:III.C.3}

As an example of the interplay between LTT/GSI effects and quantum
critical phenomena, we now consider the quantum critical behavior at
the ferromagnetic transition in quenched disordered itinerant electron
systems in $d$-dimensions. As we mentioned in the last subsection, in that 
case the two-point susceptibility $\chi^{(2)}$ has the form
\be
\chi^{(2)}(q) = c_0 + c_{d-2}\vert{\bf q}\vert^{d-2} + c_2{\bf q}^2 
    + c_{\Omega}\vert\Omega_n\vert/ {\bf q}^2 + \ldots\quad.
\label{eq:3.23}
\ee
This leads to a Gaussian part of the LGW functional
\bea
\Phi^{(2)}[M]&=&\frac{1}{2}\sum_q M(q)\,\Bigl[r
                + a_{d-2}\vert{\bf q}\vert^{d-2} + a_2\,{\bf q}^2
\nonumber\\
      &&\hskip 50pt + a_{\Omega}\vert\Omega_n\vert/\vert{\bf q}\vert^2\Bigr]
             \, M(-q)\quad.
\label{eq:3.24}
\eea
Furthermore, the field-dependent susceptibility has the form given by
Eq.\ (\ref{eq:3.22}) with $x=2$ and $y=(d-2)/2$. This implies that $u_4$
in Eq.\ (\ref{eq:3.19}), and all of the coefficients of higher-order
terms in the Landau expansion, do not exist in the limit of small
wavenumbers and frequencies. For instance, $u_4$ behaves in this limit
like
\be
u_4(\vert{\bf q}\vert\rightarrow 0,\Omega_n=0) 
     = u_4^{(d-6)}/\vert{\bf q}\vert^{d-6} + u_4^{(0)}\quad,
\label{eq:3.25}
\ee
with $u_4^{(d-6)}$ and $u_4^{(0)}$ finite numbers. 

Let us now look for a Gaussian fixed point that describes the critical
behavior. From Eq.\ (\ref{eq:3.24}) one reads off the Gaussian values
of the correlation length exponent $\nu$, the susceptibility exponents
$\gamma$ and $\eta$, and the dynamical exponent $z$ as
\bml
\label{eqs:3.26}
\be
\nu = 1/(d-2)\quad,\quad\gamma=1\quad,\quad \eta=4-d\quad,\quad z=d\quad,
\label{eq:3.26a}
\ee
for $2<d<4$, and
\be
\nu = 1/2\quad,\quad\gamma=1\quad,\quad \eta=0\quad,\quad z=4\quad,
\label{eq:3.26b}
\ee
\eml%
for $d>4$. In $d=4$ one finds mean-field exponents with logarithmic
corrections to scaling.
Power counting suggests that the non-Gaussian terms in the action are
irrelevant with respect to this fixed point, so that these exponents
represent the exact critical behavior. However, for the remaining exponents,
the order-parameter exponents $\beta$ and $\delta$, one needs to take
into account these terms, as is the case in ordinary Landau theory.
From Eq.\ (\ref{eq:3.22}) with $x=2$ we see that the average order 
parameter $m$,
which scales like $H$, scales like the wavenumber squared. The coefficient
$u_4$, Eq.\ (\ref{eq:3.25}), thus scales like $m^{-d/2}$, which suggests
an equation of state
\be
r\,m + v\,m^{d/2} + u\,m^3 = H\quad,
\label{eq:3.27}
\ee
with $u,v>0$.  From this we obtain
\bml
\label{eqs:3.28}
\be
\beta = 2/(d-2)\quad,\quad\delta = d/2\quad,
\label{eq:3.28a}
\ee
for $2<d<6$, and
\be
\beta = 1/2\quad,\quad\delta = 3\quad,
\label{eq:3.28b}
\ee
\eml%
for $d>6$. In $d=6$ there are again logarithmic corrections to scaling.
A more sophisticated analysis\cite{us_fm_dirty} confirms these results,
which illustrate the strong influence of the GSI on the critical
behavior: The exponents given in Eqs.\ (\ref{eq:3.26a}) and (\ref{eq:3.28a})
for the experimentally most interesting case $d=3$ 
are very different from the mean-field exponents one obtains if one
neglects the GSI effects.\cite{Hertz}

While the above simple power-counting arguments produce the
correct exponents of the exact critical behavior, it turns out that the
simple Gaussian fixed point presented above is marginally unstable.
This has been traced to the existence of two time scales, viz. the
critical one with dynamical exponent $z$, and the diffusive one with
dynamical exponent $2$. The existence of the latter has been obscured in
the derivation of the LGW theory, which integrates out the diffusive
modes. A more careful analysis, which keeps all of the soft 
modes explicitly and on equal footing,
shows that the actual critical fixed point is not Gaussian, but has the
same exponents as given above, with multiplicative logarithmic corrections
to scaling.\cite{us_fm_dirty_revisited} 
A convenient way to account for these corrections is to
write the critical behavior as power laws, with scale-dependent critical
exponents. For instance, the correlation length $\xi$ as a function of 
the distance $r$ from criticality
has a log-log normal correction to the simple power law,
\bml
\label{eqs:3.29}
\be
\xi \propto 1/r^{1/(d-2)}\,g(\ln (1/r))\quad,
\label{eq:3.29a}
\ee
where $g(x)$ is a function whose leading behavior at large arguments is
\be
g(x\rightarrow\infty) = 
     {\rm const.}\times e^{[\ln (c(d)\,x)]^2/2\ln (d/2)}\quad,
\label{eq:3.29b}
\ee
\eml
with $c(d)$ a dimensionality dependent constant. 
This behavior can be represented by writing
$\xi \propto r^{-1/\nu}$ with a scale dependent exponent $\nu$,
\bml
\label{eqs:3.30}
\be
1/\nu = d-2 + \ln g(b)/\ln b\quad,
\label{eq:3.30a}
\ee
with $b$ an arbitrary renormalization group length scale factor. For instance,
to translate the dependence of $b$ into an $r$-dependence, one needs to 
substitute $b = r^{-\nu}$. For the other exponents, the analysis of
Ref.\ \onlinecite{us_fm_dirty_revisited} yields
\bea
z&=&2\delta = d + \ln g(b)/\ln b\quad,
\label{eq:3.30b}\\
\eta&=&4-d - \ln g(b)/\ln b\quad,
\label{eq:3.30c}\\
\beta&=&2\nu\quad.
\label{eq:3.30d}
\eea
\eml%
Equations (\ref{eqs:3.29}) and (\ref{eqs:3.30}) are valid for $2<d<4$.
The exponent $\gamma$ has no logarithic corrections, and is $\gamma = 1$
exactly, as stated in Eqs.\ (\ref{eqs:3.26}).

The above detailed predictions have yet to be tested experimentally.

\subsubsection{Example 2: The clean quantum ferromagnetic transition}
\label{subsubsec:III.C.4}

As a second example we briefly consider the case of a clean ferromagnet. 
In this case, the coefficient $u_4$ of the quartic term in the Landau
expansion diverges like $1/\vert{\bf q}\vert^{d-3}$, and the average
order parameter $m$ scales like the wavenumber. Furthermore, as we have
mentioned in Sec.\ \ref{subsubsec:III.B.2}, the sign of the nonanalytic
term is opposite to that in the disordered case. This leads to an
equation of state 
\bml
\label{eqs:3.31}
\be
r\,m - v\,m^d + u\,m^3 = H\quad,
\label{eq:3.31a}
\ee
for general $d$, and
\be
r\,m + v\,m^3\ln m^2 + u\,m^3 = H\quad,
\label{eq:3.31b}
\ee
\eml% 
for $d=3$, with $u,v>0$.\cite{us_1st_order} 

These considerations suggest that, generically, the quantum ferromagnetic
transition of clean itinerant electrons is of first order. Indeed,
this is what is observed in MnSi,\cite{Pfleiderer_etal} 
see Fig.\ \ref{fig:8}, and in UGe$_2$,\cite{UGe_2}
both systems where the ferromagnetic transition can be triggered at very
low temperatures by applying hydrostatic pressure.
\begin{figure}[htb]
\epsfxsize=60mm
\centerline{\epsffile{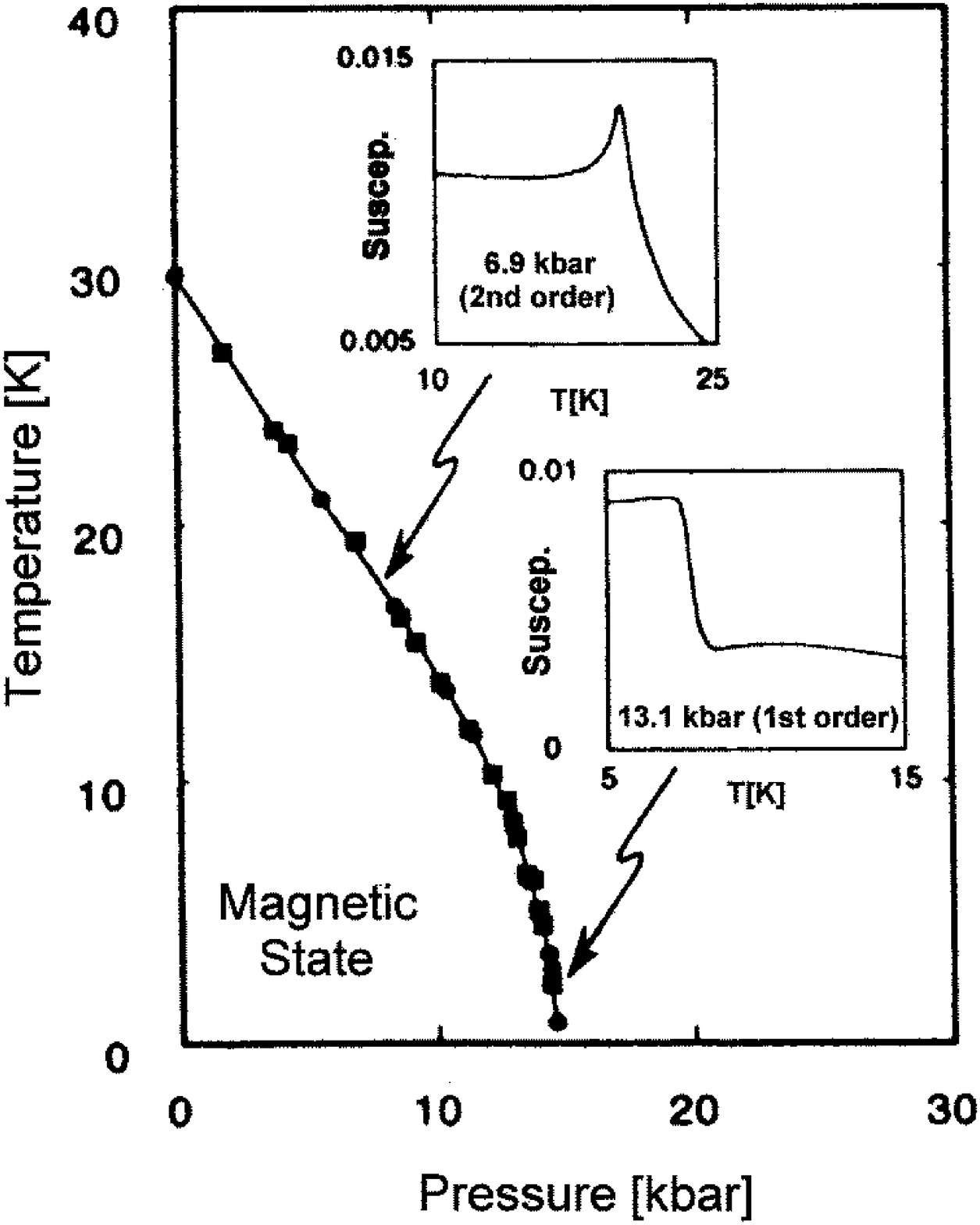}}
\vskip 5mm
\caption{Phase diagram of MnSi. The insets show the behavior of
   the susceptibility close to the transition as reported in
   Ref.\ \protect\onlinecite{Pfleiderer_etal}. 
   From Ref.\ \protect\onlinecite{us_fm_hh}.}
 \vskip 0mm
\label{fig:8}
\end{figure}
Both a nonzero temperature and nonzero disorder cut off the singularity
in Eq.\ (\ref{eq:3.31b}), and disorder, of course, induces the stronger
singularity shown in Eq.\ (\ref{eq:3.27}). For small values of both
temperature and disorder, the competition between these terms leads to
an interesting structure of the phase diagram, with tricritical points
and critical endpoints. This has been discussed in detail
in Ref.\ \onlinecite{us_1st_order}.

\subsection{Nonequilibrium effects}
\label{subsec:III.D}

Very recently, spatial correlations of density fluctuations have been
studied in noninteracting disordered electronic system that are not in
equilibrium.\cite{YoshimuraTRK}
For the model defined by Eqs.\ (\ref{eq:3.2}) and (\ref{eqs:3.6}), without
the electron-electron interaction term $S_{\rm int}$ but in the presence
of a chemical-potential gradient $\nabla\mu$, this reference
calculated an electron density-scatterer density
correlation function defined by
\bml
\label{eqs:3.32}
\be
C_1({\bf x},{\bf y}) = \left\{\langle\delta n({\bf x})\rangle\,u({\bf y})
                                      \right\}_{\rm dis}\quad,
\label{eq:3.32a}
\ee
where $\langle\ldots\rangle$ denotes a nonequilibrium thermal average.
The calculation shows that the Fourier transform of
Eq.\ (\ref{eq:3.32a}), $C_1({\bf k})$, behaves like
\be
C_1({\bf k}) = \frac{i{\bf k}\cdot\nabla\mu}{2\pi D{\bf k}^2}\quad.
\label{eq:3.32b}
\ee
\eml%
In real space, this corresponds to a decay proportional to 
$1/\vert{\bf x}\vert^{d-1}$.
This is the same result as the one in the corresponding classical Lorentz
gas, since the additional soft modes in the quantum system do not couple
to $C_1$.

It is also interesting to consider the electronic structure factor,
\be
C_2({\bf x},{\bf y}) = \left\{\langle\delta n({\bf x})\,\delta n({\bf y})
                                            \rangle\right\}_{\rm dis}\quad.
\label{eq:3.33}
\ee
Let us consider the nonequilibrium part of $C_2$. As one would expect,
the calculation yields a result that is analogous to the one in the
classical fluid, Eq.\ (\ref{eq:2.18a}), viz.
\be
C_2({\bf k}) = {N_{\rm F}\mu\tau_{\rm rel}\over 6d\pi (D{\bf k}^2)^2}\,\left[
                  25(\nabla\mu)^2 - 12({\hat{\bf k}}\cdot\nabla\mu)^2\right]
                                                                    \quad.
\label{eq:3.34}
\ee
As in the classical case, this corresponds to a decay in real space like
${\rm const.} - \vert{\bf x}\vert$ in three-dimensions.
For a classical Lorentz gas one finds instead
\be
C_2({\bf k}) \propto (\nabla\mu)^2/{\bf k}^2\quad.
\label{eq:3.35}
\ee
The weaker singularity in this case is due to the fact that the classical
model has fewer soft modes, as we discussed in Sec.\ \ref{subsubsec:III.B.1}.

\section{Conclusions}
\label{sec:IV}

In this paper we have emphasized that the soft modes which always exist in
many-body systems due to either conservation laws or broken continuous
symmetries via Goldstone's theorem, generically couple to the correlation
functions relevant for both scattering and transport experiments, and lead to
power-law correlations in space and time, in the entire phase diagram. That
is, correlation functions usually exhibit generic scale invariance.

In the classical case we have seen that these GSI effects can get amplified
either in a phase with Goldstone modes, or near continuous phase
transitions. In both cases the crucial point is that the coefficient of a
correlation function exhibiting GSI is proportional to a static
susceptibility that diverges either in an entire Goldstone phase, or at a
special critical point.

The quantum case is of particular interest because of the inherent coupling
between statics and dynamics in zero-temperature systems. Generically, this
implies that systems that exhibit GSI in the time domain
do so in space as well. This has been contrasted to classical systems, where 
the equilibrium fluid provides an example of a system that exhibits GSI in 
time correlations functions, i.e., that has LTTs, but exponentially decaying 
correlations in space. This coupling also leads to the important conclusion 
that many quantum phase transitions are in a non-mean field universality 
class governed by long-ranged interactions.

Among the many experimental consequences of the particular type of GSI
discussed here we mention, (1) transport coefficients that depend 
nonanalytically on the frequency, in both classical and quantum systems, 
(2) singularities in transport coefficients near continuous phase 
transitions, (3) enhanced, compared to equilibrium, light scattering in 
nonequilibrium systems, and, (4) quantum phase transitions that are in 
different universality classes than they would be in the absence of these 
effects. The first three of these effects have been observed. Regarding the
fourth one, the predicted first order nature of the quantum phase transition
in clean systems has been observed, but the critical behavior in disordered
systems remains to be investigated experimentally.

\acknowledgments

It is with great pleasure that we dedicate this paper to Professor J. Robert
Dorfman on the occasion of his sixty-fifth birthday. 

This work was supported
by National Science Foundation Grants Nos. DMR-98-70597 and DMR-99-75259.
The research of J.V. Sengers is supported by the Chemical Sciences,
Geosciences and Biosciences Division of the Office of Basic Energy Sciences
of the Department of Energy under Grant No. DE-FG02-95ER-14509.


\begin{references}
\b{AlderWainwright} B.J. Alder and T.E. Wainwright, Phys. Rev. A {\bf 1}:18
 (1970).
\b{DorfmanCohen} J.R. Dorfman and E.G.D. Cohen, Phys. Rev. Lett. {\bf 25}:1257 
 (1970).
\b{EHvL} M.H. Ernst, E.H.  Hauge, and J.M.J. van Leeuwen, Phys. Rev. Lett. 
 {\bf 25}:1254 (1970).
\b{WoodErpenbeck} W.W. Wood and J.J. Erpenbeck, Annu. Rev. Phys. Chem. 
 {\bf 27}:331 (1975).
\b{Dorfman} J.R. Dorfman, Physica A {\bf 106}:77 (1981).
\b{DKS} J.R. Dorfman, T.R. Kirkpatrick, and J.V. Sengers, Annu. Rev. Phys.
 Chem. {\bf 45}:213 (1994).
\b{2d_hydrodynamics} D. Forster, D.R. Nelson, and M.J. Stephen, Phys.
 Rev. A {\bf 16}:732 (1977), and references therein.
\b{BoonYip} See, e.g., J.P. Boon and S. Yip, {\it Molecular Hydrodynamics}
 (Dover, New York 1980).
\b{Stanley} H.E. Stanley, {\it Introduction to Phase Transitions and Critical
 Phenomena} (Oxford University Press, Oxford 1971).
\b{Fisher} M.E. Fisher, in {\it Advanced Course on
 Critical Phenomena}, F.W. Hahne, ed. (Springer, Berlin 1983), p.1.
\b{HohenbergHalperin} P.C. Hohenberg and B.I. Halperin, Rev. Mod. Phys. 
 {\bf 49}:435 (1977).
\b{Nagel} S. Nagel, Rev. Mod. Phys. {\bf 64}:321 (1992).
\b{Sengers_etal} J.V. Sengers, in {\it Critical Phenomena},
 NBS Miscellaneous Publ., M.S. Green and
 J.V. Sengers, eds. (U.S. Gov't Printing Office, Washington, DC, 1966).
\b{Fixman} M. Fixman, Adv. Chem. Phys. {\bf 6}:175 (1964); J. Chem. Phys.
 {\bf 47}:2808 (1967).
\b{KadanoffSwift} L.P. Kadanoff and J. Swift, Phys. Rev. {\bf 166}:89 (1968).
\b{Kawasaki} K. Kawasaki, Phys. Rev. {\bf 150}:291 (1966); 
 Ann. Phys. {\bf 61}:1 (1970).
\b{analogy_footnote} Most of these developments in the theory of quantum 
 many-body systems were put in a language that makes analogies to classical 
 LTT effects far from obvious, although early work in this area did stress 
 the connection with classical statistical mechanics, J.S. Langer and T. Neal, 
 Phys. Rev.  Lett. {\bf 16}:984 (1966). Part of our motivation in the present 
 paper is to elucidate the hidden close relation between these phenomena.
\b{gang_of_4} E. Abrahams, P.W. Anderson, D.C. Licciardello, and T.V.
 Ramakrishnan, Phys. Rev. Lett. {\bf 42}:673 (1979).
\b{NegeleOrland} See, e.g., J.W. Negele and H. Orland, 
 {\it Quantum Many-Particle Systems} (Addison-Wesley, New York 1988).
\b{WL_footnote} Historically, the term `weak localization' was first used for
 the transport properties of noninteracting disordered electrons in 
 two-dimensions. It is now often used in the broader meaning of LTT and GSI 
 phenomena in fermion systems, with or without interactions, in any
 dimensionality. For reviews, see, P.~A. Lee and T.~V. Ramakrishnan, Rev.
 Mod. Phys. {\bf 57}:287 (1985); B.L. Altshuler and A.G. Aronov, in
 {\it Electron-Electron Interactions in Disordered Systems},
 M. Pollak and A.L. Efros, eds. (North Holland, Amsterdam 1984), p.1.
\b{Anderson} P.W. Anderson, Phys. Rev. {\bf 109}:1492 (1958).
\b{Mott} N.F. Mott, {\it Metal-Insulator Transitions} (Taylor\&Francis,
 London 1990).
\b{us_R} D. Belitz and T.R. Kirkpatrick, Rev. Mod. Phys. {\bf 66}:261 (1994).
\b{Sachdev} S. Sachdev, {\it Quantum Phase Transitions} (Cambridge University
 Press, Cambridge 1999).
\b{Sondhi_etal} S.L. Sondhi, S.M. Girvin, and J.P. Carini, {\it Rev.
 Mod. Phys.} {\bf 69}:315 (1997).
\b{us_March} T.R. Kirkpatrick and D. Belitz, in {\it Electron Correlation in
 the Solid State}, N.H. March, ed. (Imperial College Press, London 1999).
\b{fluctuating_hydrodynamics} L.D. Landau and E.M. Lifshitz, 
 {\it Fluid Mechanics} (Pergamon, London 1959), ch. 17; D. Ronis, I. Procaccia,
 and J. Machta, Phys. Rev. A {\bf 22}:714 (1980).
\b{Chandrasekhar} S. Chandrasekhar, {\it Hydrodynamic and Hydromagnetic
 Stability} (Oxford University Press, Oxford 1961).
\b{Forster} D. Forster, {\it Hydrodynamic Fluctuations, Broken Symmetry, and
 Correlation Functions} (Benjamin, Reading, MA 1975).
\b{TCF_footnote} With this procedure, we are able to obtain results for the
 velocity autocorrelation function without explicitly using the properties 
 of the Langevin forces, Eqs.\ (\ref{eqs:2.2}). We do, however, need the
 various equal-time correlation functions as input. This is the same
 information that is contained in Eqs.\ (\ref{eqs:2.2}). 
\b{ZM_footnote} We note that Eq.\ (\ref{eq:2.8a}) is actually exact (see,
 e.g., Ref.\ \onlinecite{Forster}), although showing this requires a much
 more involved derivation.
\b{EHvL_2} M.H. Ernst, E.H. Hauge, and J.M.J. van Leeuwen, J. Stat. Phys.
 {\bf 15}:7 (1976); {\it ibid.} {\bf 15}:23 (1976).
\b{Kawasaki_in_DG} K. Kawasaki, in {\it Phase Transitions and Critical
 Phenomena}, vol. 5a, C. Domb and M.S. Green, eds. (Academic Press,
 New York 1976), p.165.
\b{MRT} G.F. Mazenko, S. Ramaswamy, and J. Toner, Phys. Rev. B {\bf 28}:1618
 (1983).
\b{ForsterNelsonStephen} D. Forster, D.R. Nelson, and M.J. Stephen, Ref.\ 
 \onlinecite{2d_hydrodynamics}.
\b{LSO} J. Luettner-Strathmann, J.V. Sengers, and G.A. Olchowy, J. Chem.
 Phys. {\bf 103}:7482 (1995).
\b{Imeada_etal} T. Imeada, A. Onuki, and K. Kawasaki, Progr. Theor. Phys.
 {\bf 71}:16 (1984).
\b{KCD} T.R. Kirkpatrick, E.G.D. Cohen, and J.R. Dorfman, Phys. Rev. A 
 {\bf 26}:950 (1982), {\it ibid.} {\bf 26}:995 (1982).
\b{SchmitzCohen} R. Schmitz and E.G.D. Cohen, J. Stat. Phys. {\bf 40}:431 
 (1985)
\b{Ortiz_etal} J.M. Ortiz de Zarate, R. Perez Cordon, and J.V. Sengers, 
 Physics A {\bf 291}:113 (2001).
\b{LawSengers} B.M. Law and J.V. Sengers, J. Stat. Phys. {\bf 57}:531 (1989).
\b{LSGS} W.B. Li, P.N. Segr{\`e}, R.W. Gammon, and J.V. Sengers, Physica
 A {\bf 204}:399 (1994).
\b{LawNieuwoudt} B.~M. Law and J.~C. Nieuwoudt, Phys. Rev. A {\bf 40}:3880
 (1989).
\b{SengersOrtiz} J.V. Sengers and J.M. Ortiz de Z{\'a}rate, in {\it Thermal
 Nonequilibrium Phenomena in Fluid Mixtures}, W. K{\"o}hler and W. Wiegand,
 eds. (Springer, Berlin 2001).
\b{LZSGO} W.B. Li, K.J. Zhang, J.V. Sengers, R.W. Gammon, and J.M. Ortiz de
 Z{\'a}rate, Phys. Rev. Lett. {\bf 81}:5580 (1998); J. Chem. Phys. 
 {\bf 112}:9139 (2000).
\b{NESS} T. Imeada, A. Onuki, and K. Kawasaki, Ref.\ \onlinecite{Imeada_etal};
 B. Schmittmann and R.K.P. Zia, in {\it Phase Transitions and Critical
 Phenomena}, Vol. 17, C. Domb and J.L. Lebowitz, eds. (Academic Press, New
 York, 1995).
\b{us_fermionsII} See, e.g., D. Belitz, F. Evers, and T.R. Kirkpatrick,
 Phys. Rev. {\bf 58}:9710 (1998).
\b{FL} See, e.g., G. Baym and C. Pethick, {\it Landau 
 Fermi-Liquid Theory} (Wiley, New York 1991).
\b{Wegner} F. Wegner, Z. Phys. B {\bf 35}:207 (1979).
\b{WegnerSchaefer} L. Sch{\"a}fer and F. Wegner, Z. Phys. B {\bf 38}:113 
 (1980).
\b{us_fermions} D. Belitz and T.R. Kirkpatrick, Phys. Rev. B {\bf 56}:6513
 (1997).
\b{GLK} L.P. Gorkov, A.I. Larkin, and D.E. Khmelnitskii, Pis'ma Zh. Eksp. Teor.
 Fiz. {\bf 30}:248 (1979) [JETP Lett. {\bf 30}:228 (1979)].
\b{AAL} B.L. Altshuler, A.G. Aronov, and P.A. Lee, Phys. Rev. Lett. 
 {\bf 44}:1288 (1980).
\b{Hauge} See, e.g., E.H. Hauge in {\it Transport Phenomena}, Lecture Notes in
 Physics No. 31, G. Kirczenow and J. Marro, eds. (Springer, New York 1974),
 p.337.
\b{Ernst_et_al} M.H. Ernst, J. Machta, H. van Beijeren, and J.R. Dorfman,
 J. Stat. Phys. {\bf 34}:477 (1984).
\b{Dai_etal} P. Dai, Y. Zhang, and M.P. Sarachik, Phys. Rev. B {\bf 45}:3984
 (1992).
\b{DolanOsheroff} G.J. Dolan and D.D. Osheroff, Phys. Rev. Lett. {\bf 43}:721 
 (1979).
\b{us_ernst} T.R. Kirkpatrick and D. Belitz, J. Stat. Phys. {\bf 87}:1307
 (1997).
\b{2d_MIT_footnote} For noninteracting electrons, there is consensus that this
 statement is true. Until recently, it was also believed to be true for
 interacting electrons, but experiments showing an apparent metal-insulator
 transition in two-dimensional systems have cast doubt on this, see
 E. Abrahams, S.V. Kravchenko, and M.P. Sarachik, Rev. Mod. Phys. {\bf 73}:251
 (2001). The correct explanation of these observations is still being debated.
\b{us_fm_dirty} T.R. Kirkpatrick and D. Belitz, Phys. Rev. B {\bf 53}:14364
 (1996).
\b{us_OP} D. Belitz, T.R. Kirkpatrick, and T. Vojta, cond-mat/0109547.
\b{us_chi_s} D. Belitz, T.R. Kirkpatrick, and T. Vojta, Phys. Rev. B 
 {\bf 55}:9452 (1997).
\b{Millis} G.Y. Chitov and A.J. Millis, cond-mat/0103155.
\b{Pfleiderer_etal} C. Pfleiderer, G.J. McMullan, S.R. Julian, and 
 G.G. Lonzarich, Phys. Rev. {\bf 55}:8330 (1997).
\b{UGe_2} S.S. Saxena et al., Nature {\bf 406}:587 (2000); A. Huxley
 et al., Phys. Rev. B {\bf 63}:144519 (2001).
\b{ZrZn_2} C. Pfleiderer et al., Nature {\bf 412}:58 (2001).
\b{WilsonKogut} See, K.G. Wilson and J. Kogut, Phys. Rep. {\bf 12}:75
 (1974) for the general philosophy underlying this approach. For the
 application to quantum phase transitions, see J. Hertz, Ref.\ 
 \onlinecite{Hertz}, and references therein.
\b{Hertz} J. Hertz, Phys. Rev. B {\bf 14}:1165 (1976).
\b{HS} J. Hubbard, Phys. Rev. Lett. {\bf 3}:77 (1959); R.L. Stratonovich,
 Dok. Akad. Nauk. SSSR {\bf 115}:1907 (1957) [Sov. Phys. Dokl. {\bf 2}:416 
 (1957)].
\b{us_fm_dirty_revisited} D. Belitz, T.R. Kirkpatrick, M.T. Mercaldo, and
 S.L. Sessions, Phys. Rev. B {\bf 63}:174427 (2001); {\it ibid.} 
 {\bf 63}:174428 (2001).
\b{us_1st_order} D. Belitz, T.R. Kirkpatrick, and T. Vojta, Phys. Rev. Lett.
 {\bf 82}:4707 (1999).
\b{us_fm_hh} T. Vojta, D. Belitz, T.R. Kirkpatrick, and R. Narayanan, Ann.
 Phys. (Leipzig) {\bf 8}:593 (1999).
\b{YoshimuraTRK} M. Yoshimura and T.R. Kirkpatrick, Phys. Rev. B {\bf 54}:7109 
 (1996).
\end{references}
\end{document}